\documentclass[oldversion]{aa}  

\usepackage{graphicx}
\usepackage{txfonts}
\usepackage{float}
\usepackage[latin1]{inputenc}
\usepackage{supertabular}
\usepackage[normalem]{ulem}

\usepackage{natbib}
\bibpunct{(}{)}{;}{a}{}{,}

\newcommand{\FAC}{HE~1327$-$2326} 
\newcommand{\CBB}{HE~0107$-$5240} 
\newcommand{\cd}{CD~$-38^{\circ}\,245$}            
\newcommand{\rb}[1]{\raisebox{1.5ex}[-1.5ex]{#1}}

\begin{document}

\title{The stellar content of the Hamburg/ESO survey\thanks{Based on
    observations collected at Las Campanas Observatory, Palomar Observatory,
    Siding Spring Observatory, and the European Southern Observatory (Proposal
    IDs 69.D-0130, 170.D-0010, 073.D-0555, and 081.D-0596).}}

\subtitle{V. The metallicity distribution function of the Galactic halo}

\author{
  T. Sch\"orck\inst{1}            \and
  N. Christlieb\inst{2,3}         \and
  J.G. Cohen\inst{4}              \and
  T.C. Beers\inst{5}              \and
  S. Shectman\inst{6}             \and
  I. Thompson\inst{6}             \and
  A. McWilliam\inst{6}            \and
  M.S. Bessell\inst{7}            \and
  J.E. Norris\inst{7}             \and
  J. Mel{\'e}ndez\inst{8}         \and
  S. Solange Ram\'{\i}rez\inst{9} \and
  D. Haynes\inst{10}              \and
  P. Cass\inst{10}                \and
  M. Hartley\inst{10}             \and
  K. Russell\inst{10}             \and
  F. Watson\inst{10}              \and
  F.-J. Zickgraf\inst{1}          \and
  B. Behnke\inst{11}              \and
  C. Fechner\inst{12}             \and
  B. Fuhrmeister\inst{1}          \and
  P.S. Barklem\inst{3}            \and
  B. Edvardsson\inst{3}           \and
  A. Frebel\inst{13}              \and
  L. Wisotzki\inst{14}            \and
  D. Reimers\inst{1}
}

\offprints{N. Christlieb,\\ \email{N.Christlieb@lsw.uni-heidelberg.de}}

\institute{
     Hamburger Sternwarte, Universit\"at Hamburg, Gojenbergsweg 112,         
     D-21029 Hamburg, Germany;
     \email{tschoerck/fzickgraf/dreimers@hs.uni-hamburg.de}
\and Zentrum f\"ur Astronomie der Universit\"t Heidelberg, Landessternwarte,  
     K\"onigstuhl 12, 69117 Heidelberg, Germany;
     \email{N.Christlieb@lsw.uni-heidelberg.de}
\and Department of Physics and Astronomy, Uppsala University, Box 515,       
     SE-75120 Uppsala, Sweden; \email{barklem/be@fysast.uu.se}
\and Palomar Observatory, Mail Code 105-24, California Institute of          
     Technology, Pasadena, CA 91125, USA; \email{jlc@astro.caltech.edu}
\and Department of Physics and Astronomy, and JINA: Joint Institute for      
     Nuclear Astrophysics, Michigan State University, E. Lansing, MI 48824, 
     USA; \email{beers@pa.msu.edu}
\and Carnegie Observatories of Washington, 813 Santa Barbara Street,         
     Pasadena, CA 91101, USA; \email{shec/ian/andy@ociw.edu}
\and Research School of Astronomy and Astrophysics, Australian National      
     University, Cotter Road, Weston, ACT 2611, Australia;
     \email{bessell/jen@mso.anu.edu.au}
\and Centro de Astrofisica da Univ. Porto, Rua das Estrelas, 4150-762 Porto, 
     Portugal; \email{jorge@astro.up.pt}
\and IPAC, Mail Code 100-22, California Institute of Technology, Pasadena,   
     CA 91125, USA; \email{solange@ipac.caltech.edu}
\and Anglo-Australian Observatory; PO Box 296, Epping, NSW 1710, Australia;  
     \email{dmj/cpc/mh/ksr/fgw@aao.gov.au}
\and MPI for Gravitational Physics, Albert-Einstein-Institute, Am 
     M\"hlenberg, D-14476 Golm, Germany;
     \email{Berit.Behnke@aei.mpg.de}
\and Universit\"at Potsdam, Institut f\"ur Physik und Astronomie,            
     Karl-Liebknecht-Stra{\ss}e 24/25, D-14476 Potsdam, Germany;
     \email{cfech@astro.physik.uni-potsdam.de}
\and McDonald Observatory, The University of Texas at Austin, 1 University   
     Station, C1400, Austin, TX~78712-0259; 
     \email{anna@astro.as.utexas.edu}
\and Astrophysical Institute Postsdam, An der Sternwarte 16,                 
     D-14482 Potsdam, Germany; \email{lutz@aip.de}
}

\date{Received 8 September 2008 / Accepted 11 September 2009}

\abstract{
  
  We determine the metallicity distribution function (MDF) of the Galactic
  halo by means of a sample of 1638 metal-poor stars selected from the
  Hamburg/ESO objective-prism survey (HES). The sample was corrected for minor
  biases introduced by the strategy for spectroscopic follow-up observations
  of the metal-poor candidates, namely ``best and brightest stars first''.
  Comparison of the metallicities [Fe/H] of the stars determined from
  moderate-resolution (i.e., $R\sim 2000$) follow-up spectra with results
  derived from abundance analyses based on high-resolution spectra (i.e.,
  $R>20,000$) shows that the [Fe/H] estimates used for the determination of
  the halo MDF are accurate to within 0.3\,dex, once highly C-rich stars are
  eliminated. We determined the selection function of the HES, which must be
  taken into account for a proper comparison between the HES MDF with MDFs of
  other stellar populations or those predicted by models of Galactic chemical
  evolution. The latter show a reasonable agreement with the overall shape of
  the HES MDF for $\mbox{[Fe/H]} > -3.6$, but only a model of Salvadori et al.
  (2007) with a critical metallicity for low-mass star formation of
  $Z_{\mathrm cr}=10^{-3.4}\,Z_{\odot}$ reproduces the sharp drop at
  $\mbox{[Fe/H]} \sim -3.6$ present in the HES MDF.  Although currently about
  ten stars at $\mbox{[Fe/H]} < -3.6$ are known, the evidence for the
  existence of a tail of the halo MDF extending to $\mbox{[Fe/H]} \sim -5.5$
  is weak from the sample considered in this paper, because it only includes
  two stars $\mbox{[Fe/H]} < -3.6$.  Therefore, a comparison with theoretical
  models has to await larger statistically complete and unbiased samples.  A
  comparison of the MDF of Galactic globular clusters and of dSph satellites
  to the Galaxy shows qualitative agreement with the halo MDF, derived from
  the HES, once the selection function of the latter is included. However,
  statistical tests show that the differences between these are still highly
  significant.

}

\titlerunning{The halo metallicity distribution function}
\authorrunning{Sch{\"o}rck et al.}

\maketitle

\section{Introduction}\label{Sect:Intro}

One of the key observables for constraining models of the formation and
chemical evolution of the Galaxy is the Metallicity Distribution Function
(MDF) of the constituent stars of its various components (bulge, disk, halo).
The MDF provides critical information on the enrichment history of those
components with heavy elements. In the case of the halo, early enrichment may
have been provided by the very first generations of massive stars, formed from
material of primordial composition shortly after the Big Bang (i.e.,
Population III stars).

Models of Galactic chemical evolution need to be compared to an accurate (and
precise) observed halo MDF to test their predictions, to constrain their
various parameters (such as the effective yield, the star-formation rate and
the IMF), and in order to obtain information on the properties of Population
III stars that are responsible for the earliest enrichment. This is
particularly important for the lowest metallicity tail of the MDF, which
provides invaluable information on the earliest enrichment phases
\citep{Prantzos:2003}; for instance, it has been suggested that a minimum
level of enrichment is required to form low-mass stars. This critical
metallicity ranges between $10^{-4}\mathrm{Z}_{\odot}$
\citep{Omukai:2000,Brommetal:2001,Bromm/Loeb:2003,Umeda/Nomoto:2003,Santoro/Shull:2006,Frebeletal:2007c}
and $10^{-6}\mathrm{Z}_{\odot}$, the latter being applicable when dust grains
are present
\citep{Schneideretal:2002,Schneideretal:2003,Schneideretal:2006,Omukaietal:2005,Tsuribe/Omukai:2006,Clarketal:2008}.

The precision of a derived halo MDF increases directly with the total number
of observed metal-poor halo stars. Selection of such stars without the
introduction of a kinematic bias (e.g., from among high proper motion stars)
makes them of particular utility for examination of the relationships between
the chemistry and kinematics of the halo. Early determinations of the halo MDF
were based on small samples of globular clusters (\citealt{Hartwick:1976};
$N=60$), or a mixture of halo subdwarfs and globular clusters
(\citealt{Bond:1981}; $N=90$ and $N=31$, respectively).  Problems with these
samples arise not only from their small sizes, but also their inaccurate
metallicities. Later studies employed significantly larger samples with
spectroscopically-determined stellar abundances. For example,
\citet{Ryan/Norris:1991} used a sample of 372 kinematically-selected halo
stars. \citet{Ryan/Norris:1991} and \citet{Carneyetal:1996} showed that the
MDF peaks at a metallicity of $\mathrm{[Fe/H]} = -1.6$ with wings from
$\mathrm{[Fe/H]} = -3.0$ to solar abundances.

The HK survey \citep{BPSI,BPSII,TimTSS}, originated by Preston and Shectman,
and greatly extended by Beers to include several hundred additional
objective-prism plates, was, until the advent of the Hamburg/ESO Survey (HES;
see below), the primary source of metal-poor candidates suitable for
consideration of the halo MDF. With the assistance of numerous colleagues,
medium-resolution spectroscopy of over 10,000 HK-survey stars was obtained,
using 1.5--4\,m class telescopes, over the past two decades. This led to the
identification of thousands of stars with $\mathrm{[Fe/H]} < -2.0$, as well as
significant numbers of stars with $\mathrm{[Fe/H]} < -3.0$.

\begin{figure}[htbp]
  \centering
  \includegraphics[clip=true,bb=75 199 372 770,width=8.8cm]{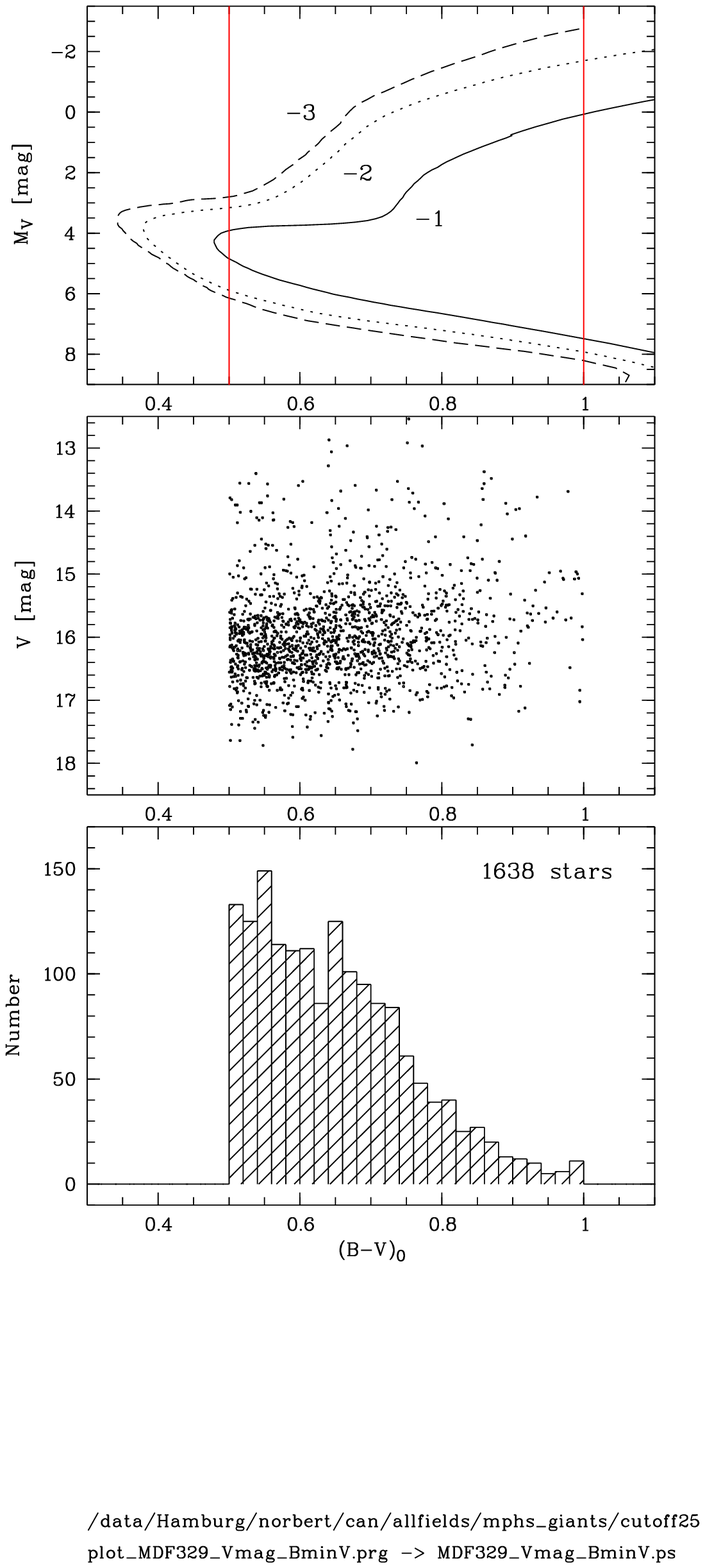}
  \caption{\label{Fig:VmagBminV} Upper panel: Isochrones for an age of 12\,Gyr 
    and metallicities of $\mbox{[Fe/H]}=-1$, $-2$, and $-3$
    \citep{Kimetal:2002}, and chosen colour cuts (see text for details);
    middle panel: $V$ magnitude distribution of the HES sample from which we
    construct the halo MDF; lower panel: $(B-V)_0$ distribution.}
\end{figure}

Another wide-angle spectroscopic survey is the HES. It was originally
conceived as a survey for bright quasars
\citep{Reimers:1990,hespaperI,hespaperIII}; however, its data quality is
sufficient to not only efficiently select quasars with redshifts of up to $z =
3.2$, but also various types of stellar objects, including metal-poor stars
\citep{Christliebetal:2008}. So far, several hundred new stars at
$\mathrm{[Fe/H]}<-3.0$ have been identified, including three stars that were
confirmed by high-resolution spectroscopy to have $\mathrm{[Fe/H]} < -4.0$:
{\FAC} ($\mathrm{[Fe/H]} = -5.4$;
\citealt{Frebeletal:2005,Aokietal:2006,Frebeletal:2006a}); {\CBB}
($\mathrm{[Fe/H]} = -5.3$;
\citealt{HE0107_Nature,HE0107_ApJ,Besselletal:2004}); and HE~0557$-$4840
($\mathrm{[Fe/H]} = -4.8$; \citealt{Norrisetal:2007}). It is perhaps of
interest that the HK survey has not (to date) yielded any stars with
$\mathrm{[Fe/H]} < -4.0$ confirmed by high-resolution spectroscopy; this may
be related to the fact that the HK survey reaches apparent magnitudes that are
brighter than the HES, and as a result is dominated more than the HES by
inner-halo stars.

The Sloan Digital Sky Survey (SDSS; \citealt{Gunnetal:1998},
\citealt{Yorketal:2000}), and in particular the Sloan Extension for Galactic
Understanding and Exploration (SEGUE), has provided even larger samples of
halo stars, as discussed by \citet{Carolloetal:2007} and
\citet{Ivezicetal:2008}. The former emphasize the division of the halo into
two structural components, an inner region with $R < 10$--$15$\,kpc, and an
outer region beyond that radius. These two components differ in stellar
metallicities, stellar orbits, and spatial density profiles. As we discuss in
Sect.~\ref{Sect:Selection} below, the HES sample is dominated by inner-halo
stars. We note that we hereafter refer to the inner halo as ``the halo'',
unless indicated otherwise.

In spite of the very large sample of $\sim 20,000$ stars used by Carollo et
al., their coverage of the regime of very low metallicity is limited.
According to their supplemental Fig.~4, they find only 3 stars with
$\mbox{[Fe/H]} < -3.0$ in their ``local sample'' of 10,123 stars. The main
reason for this is that the stars of their sample were not selected to be
metal-poor, but for the purpose of spectrophotometric and telluric calibration
of the SDSS spectra.

Recent high-resolution spectroscopic follow-up of stars from the Carollo et al.
sample (W. Aoki, priv. comm.) has indicated that the current version of the
SEGUE Stellar Parameter Pipeline (SSPP; see
\citealt{Leeetal:2008a,Leeetal:2008b,AllendePrietoetal:2008}) is somewhat
conservative in the assignment of stellar metallicity estimates, in the sense
that stars assigned $\mathrm{[Fe/H]} < -2.7$ by the SSPP are in reality more
metal-deficient, on average, by on the order of 0.3\,dex. A recent examination
of the numbers of stars from the SDSS/SEGUE survey, taking into account this
offset, suggests that up to several hundred stars with $\mathrm{[Fe/H]} <
-3.0$ are in fact present in the current SDSS sample of stars (including other
categories of targets than just the calibration stars).

\citet{Ivezicetal:2008} focus on the comparison between the inner halo and the
disk. Since they rely on abundances determined from photometry, they cannot
reliably determine metallicities of stars at $\mbox{[Fe/H]} < -2$.
Nevertheless, the metallicity map of some 2.5 million stars with photometric
metallicies shown in Fig.~8 of Ivezic et al. indicates that there exist very
large numbers of stars in SDSS consistent with $\mathrm{[Fe/H]} < -2.0$.
Follow-up spectroscopy is, at present, only available for a subset of them.
Beers et al. (in preparation) discuss the MDF of the lowest metallicity stars
found in SDSS/SEGUE. The total number of stars with $\mathrm{[Fe/H]} < -2.0$,
based on medium-resolution SDSS spectroscopy, is over 25,000 (i.e., five times
the number discovered by the combination of the HK and HES).

This paper continues our series on the stellar content of the HES
(\citealt{HESStarsI}, Paper~I; \citealt{HESStarsII}, Paper~II;
\citealt{HESFHBA}, Paper~III; \citealt{Christliebetal:2008}, Paper~IV). We are
mainly concerned with the low-metallicity tail of the halo MDF, which is
constructed from a sample of 1638 metal-poor stars selected in the HES by
quantitative criteria (Sect.~\ref{Sect:Selection}). The follow-up observations
and determination of the metallicities are described in
Sect.~\ref{Sect:FollowUp}. In Sect.~\ref{Sect:MDFconstruction} we detail how
the MDF was constructed. We discuss the shape of the halo MDF in Sect.
\ref{Sect:MDFshape}. Comparisons of the observed MDF with MDFs predicted by
models of Galactic chemical evolution are presented in
Sect.~\ref{Sect:TheoryObservations}, and a comparison with the MDFs of the
Galactic globular cluster system and dwarf spheroidal galaxies is presented in
Sect.~\ref{Sect:GCdSph}. The results are discussed in
Sect.~\ref{Sect:Conclusions}.


\begin{table}[htbp]
 \centering
 \caption{Number of stars in each candidate class in the total sample of
   candidates, number of observed candidates, and number of accepted
   candidates after removal of emission line objects, ``peculiar'' objects
   (e.g., objects with continuous spectra) and all stars with a G-band index
   $\mathrm{GP} > 6$\,{\AA}. In the last column, we list the scaling factors 
   applied to the [Fe/H] histograms for each candidate class during the 
   construction of the MDF (see Section~\ref{Sect:MDFconstruction}).}
 \label{Tab:CandidateStatistics}
  \begin{tabular}{lrrrr}\hline\hline
    \rule{0ex}{2.3ex} & \multicolumn{3}{c}{Number of stars} &  \\\cline{2-4}
    \rb{Class}        & \multicolumn{1}{c}{All} & \multicolumn{1}{c}{Observed}  
    \rule{0ex}{2.3ex} & \multicolumn{1}{c}{Accepted} & \rb{Factor}\\\hline
    \texttt{mpca} &  201 &  123 &   105 & \rule{0ex}{2.3ex} 1.63\\ 
    \texttt{unid} &  231 &  208 &   192 &                   1.11\\
    \texttt{mpcb} & 2006 & 1008 &   940 &                   1.99\\
    \texttt{mpcc} & 1275 &  432 &   401 &                   2.95\\\hline
    Sum           & 3713 & 1771 &  1638 & \rule{0ex}{2.3ex}\\\hline
  \end{tabular}
\end{table}

\section{The metal-poor star sample}\label{Sect:Selection}

One of the main advantages of the HES for determining the halo MDF is that the
selection of candidate metal-poor stars was done with quantitative criteria.
Hence, the selection is well-understood, and possible selection biases can be
quantified and corrected for during the construction of the MDF. Futhermore,
the selection is purely spectroscopic, so it does not introduce any kinematic
biases.

\begin{figure}[htbp]
  \centering
  \includegraphics[clip=true,bb=72 568 372 770,width=8.0cm]{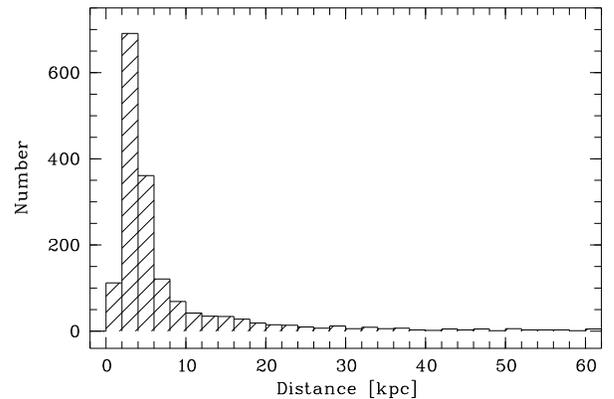}
  \caption{\label{Fig:DistHisto} Distance distribution of the HES sample.
    The sample is dominated by stars at distances of less than $\sim 5$\,kpc
    from the Sun; a few cool giants are located at distances of up to  
    $\sim 50$\,kpc.}
\end{figure}

\begin{table*}[htbp]
 \centering
 \caption{Follow-up observations of the candidate metal-poor stars.}
 \label{Tab:FollowUpObservations}
  \begin{tabular}{lllr}\hline\hline
    Telescope(s)    & Instrument(s)         & Observers & \rule{0ex}{2.3ex} $N_{\mathrm{stars}}$\\\hline
    Magellan 1\&2   & B\&C\rule{0ex}{2.3ex} & Shectman, McWilliam, Thompson          & 553     \\[0.5ex]
                    &                       & Bessell, Norris, Edvardsson,           &         \\[-0.3ex]
    \rb{SSO 2.3\,m} & \rb{DBS}              & Behnke, Christlieb, Frebel             & \rb{339}\\[0.5ex]
    Palomar 200''   & DS                    & Cohen, Ram\'{\i}rez, Melendez          & 323     \\[0.5ex]
    UK Schmidt      & 6dF                   & Haynes, Cass, Hartley, Russell, Watson & 283     \\[0.5ex]
                    &                       & Fechner, Zickgraf, Barklem,            &         \\[-0.3ex]
    \rb{ESO 3.6\,m} & \rb{EFOSC2}           & Fuhrmeister, Christlieb                & \rb{140}\\\hline
    \multicolumn{3}{l}{Total}\rule{0ex}{2.3ex}                                       & 1638\\\hline
  \end{tabular}
\end{table*}

\begin{figure}[htbp]
  \centering
  \includegraphics[clip=true,bb=54 40 422 565,width=8.0cm]{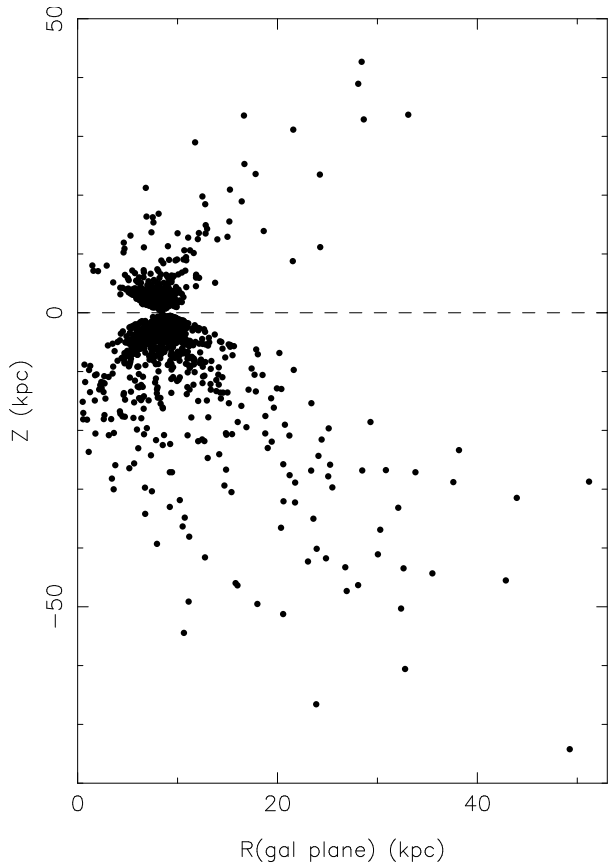}
  \caption{\label{Fig:GalHeight} Spatial distribution of the HES sample.
    $R(\mbox{gal. plane})$ is the distance from the Galactic center projected 
    onto the Galactic plane.}
\end{figure}

The selection of candidates in the HES is described in Paper~IV. For the
sample used in this study, we employed only the KP/$(B-V)_0$ selection; i.e.,
a star is selected as a metal-poor candidate if its KP index of the
\ion{Ca}{ii}~K line, as measured in its digital HES objective-prism spectrum,
is smaller than the KP index predicted for a star of $\mbox{[Fe/H]}=-2.5$ and
the same $(B-V)_0$ colour (see Fig.~5 of Paper~IV). This cutoff was chosen
because it results in a good compromise between completeness at
$\mbox{[Fe/H]}<-3.0$, the region in [Fe/H] we are mainly interested in
(because it corresponds to the earliest phases of Galactic chemical
evolution), and achieving a selection that efficiently rejects stars at higher
metallicity.  In addition to the KP index, the $B-V$ colours are measured in
the HES spectra as well (see Paper~IV for details), and then are corrected for
reddening using the maps of \citet{Schlegeletal:1998}. We restrict the sample
to the colour range $0.5 < (B-V)_0 < 1.0$, because the follow-up observations
of stars bluer than $(B-V)_0 = 0.5$ have not yet reached a sufficient level of
completeness, and for stars redder than $(B-V)_0 = 1.0$, the accuracy of the
determination of [Fe/H] from moderate-resolution follow-up spectra is limited
due to the lack of calibration stars and the weakness of the H$\delta$ line,
which is used as a temperature indicator. The $V$ magnitude and $(B-V)_0$
distribution of our sample together with isochrones for an age of 12\,Gyr and
different metallicities is shown in Fig.~\ref{Fig:VmagBminV}. The $V$
magnitudes as well as the $(B-V)_0$ colours are from the HES.

The selection was applied to all spectra of unsaturated point sources
extracted on 329 (out of 379) HES plates, covering a nominal area of $\sim
7700$\,deg$^2$ of the southern high galactic latitude sky. The candidates were
visually inspected and assigned to the classes \texttt{mpca}, \texttt{unid},
\texttt{mpcb}, and \texttt{mpcc}. As described in Paper~IV, the classification
is based on the appearance of the \ion{Ca}{ii}~K line in the digital HES
spectra. Candidates of class \texttt{mpca} are the best in terms of the
success rate of finding stars at $\mathrm{[Fe/H]}<-2.5$ (see
Fig.~\ref{Fig:HES_MDF_byclass}), since no \ion{Ca}{ii}~K line could be seen in
the HES spectrum, while the candidates of class \texttt{mpcc} are the worst,
because a strong Ca~K line could clearly be seen. However, the Ca~K line is
still strong in cool, moderately metal-poor (i.e., $\mbox{[Fe/H]}\sim -2.0$)
giants, therefore the line is expected to be detected in the HES spectra of
such stars. For statistical studies such as the determination of the halo MDF
it is therefore necessary to obtain follow-up spectroscopy also of the
\texttt{mpcc} candidates, because otherwise a color-related bias would be
introduced. Furthermore, the assignment of classes to the candidates is
subjective, and therefore it would be impossible to determine the selection
function of the HES if only a subset of the candidates selected by
quantitative criteria would be considered for the construction of the MDF.

\begin{figure*}[htbp]
  \centering
  \includegraphics[clip=true,bb=49 30 286 629,width=8cm]{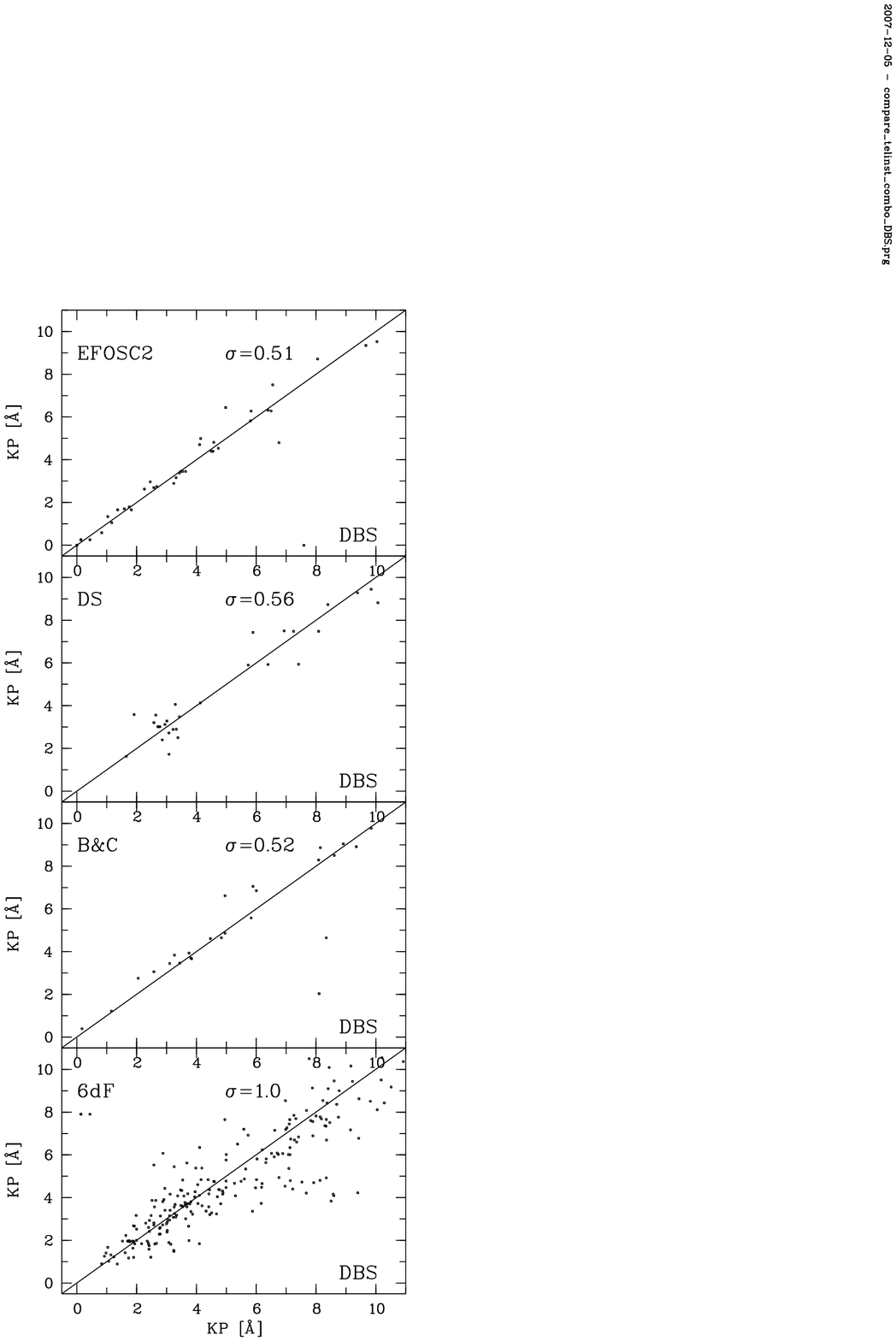}
  \includegraphics[clip=true,bb=49 30 286 629,width=8cm]{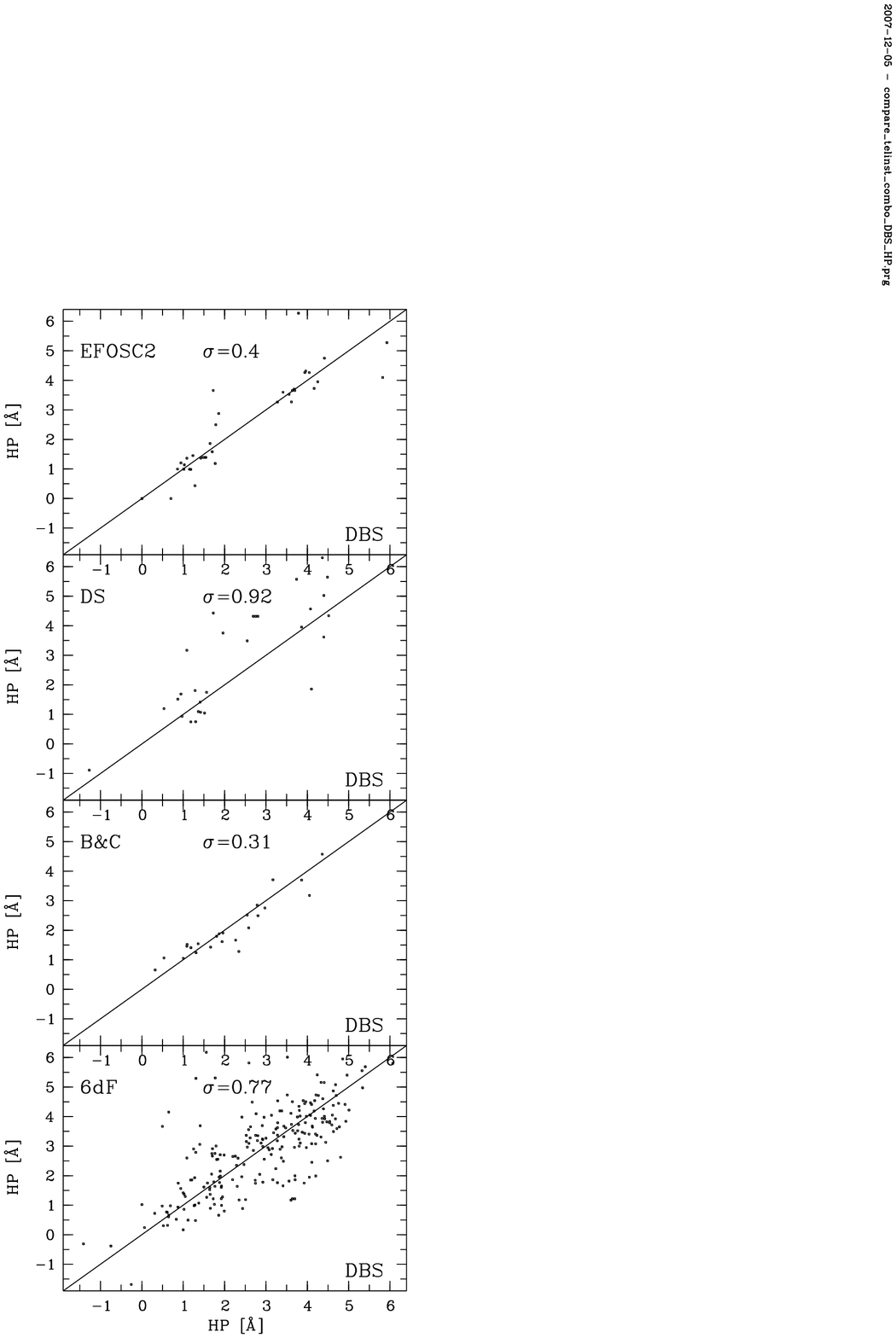}
  \caption{\label{Fig:KPHP2test} Pairs of KP and HP2 measurements for the
    same star in spectra obtained with different telescope/instrument
    combinations. Note that some of the estimates of $\sigma$ displayed in the
    panels are influenced by a number of outliers; i.e., robust estimates
    would yield smaller values.}
\end{figure*}

The result of the visual inspection are 3792 accepted candidates, of which 79
are present on multiple plate quarters or plates; the number of unique
candidates is 3713. The number of candidates in each of the aforementioned
classes is listed in Tab.~\ref{Tab:CandidateStatistics}.  Only about half of
the 3713 candidates are part of the sample presented in Tab.~A.1 of Paper~IV.
This is because slightly improved sky background and spectrum extraction
algorithms were used in the final reduction of the HES, from which the sample
of Paper~IV was drawn. While minor changes of the reduction algorithms can
have a large effect on the measurement of the KP index in individual spectra,
because the \ion{Ca}{ii}~K line is covered by only four pixels of the HES
spectra, we verified (HES plate by HES plate) that there are no systematic
differences between the KP indices measured in spectra reduced with the older
extraction algorithms and the spectra to which the selection described in
Paper~IV was applied. Therefore, there should not be any \emph{statistical}
differences between the HES metal-poor sample presented in Paper~IV and the
sample used in this paper. We decided to construct the halo MDF from an older
sample because the spectroscopic follow-up observations of that sample is more
advanced, resulting in a considerably larger sample size.

We determine distances to each of the sample stars using the [Fe/H] for each
star and a set of isochrones similar to those shown in the upper panel of
Fig.~\ref{Fig:VmagBminV}.  Assuming that all the sample stars are at or above
the main-sequence turnoff, we obtain the distance distribution shown in
Fig.~\ref{Fig:DistHisto}, and the spatial distribution shown in
Fig.~\ref{Fig:GalHeight}. The cooler giants in our sample reach distances from
the Galactic plane well beyond $|Z| = 15$\,kpc. However, the sample is clearly
dominated by inner-halo stars. There is a hint that the outer-halo stars with
$|Z| > 15$\,kpc have a higher fraction of extremely metal-poor stars than do
those of the inner halo with $5 < |Z| < 15$\,kpc, but given the wide range in
metallicity we see throughout the halo, our sample is too small to determine
the MDFs of the inner and outer halo separately with confidence.


\section{Determination of metallicities}\label{Sect:FollowUp}

For 1771 of the 3713 unique candidates, moderate-resolution spectroscopy was
obtained with various telescope/instrument combinations (see
Tab.~\ref{Tab:FollowUpObservations}). The candidates were mostly observed in
programs aiming at the identification of targets for high-resolution
spectroscopy of the most metal-poor stars. Hence, the observing strategy
adopted for the follow-up observations was to observe the brightest and best
candidates (i.e., candidate classes \texttt{mpca} and \texttt{unid}) first.

\begin{figure}[htbp]
  \centering
  \includegraphics[clip=true,bb=75 382 373 771,width=8.8cm]{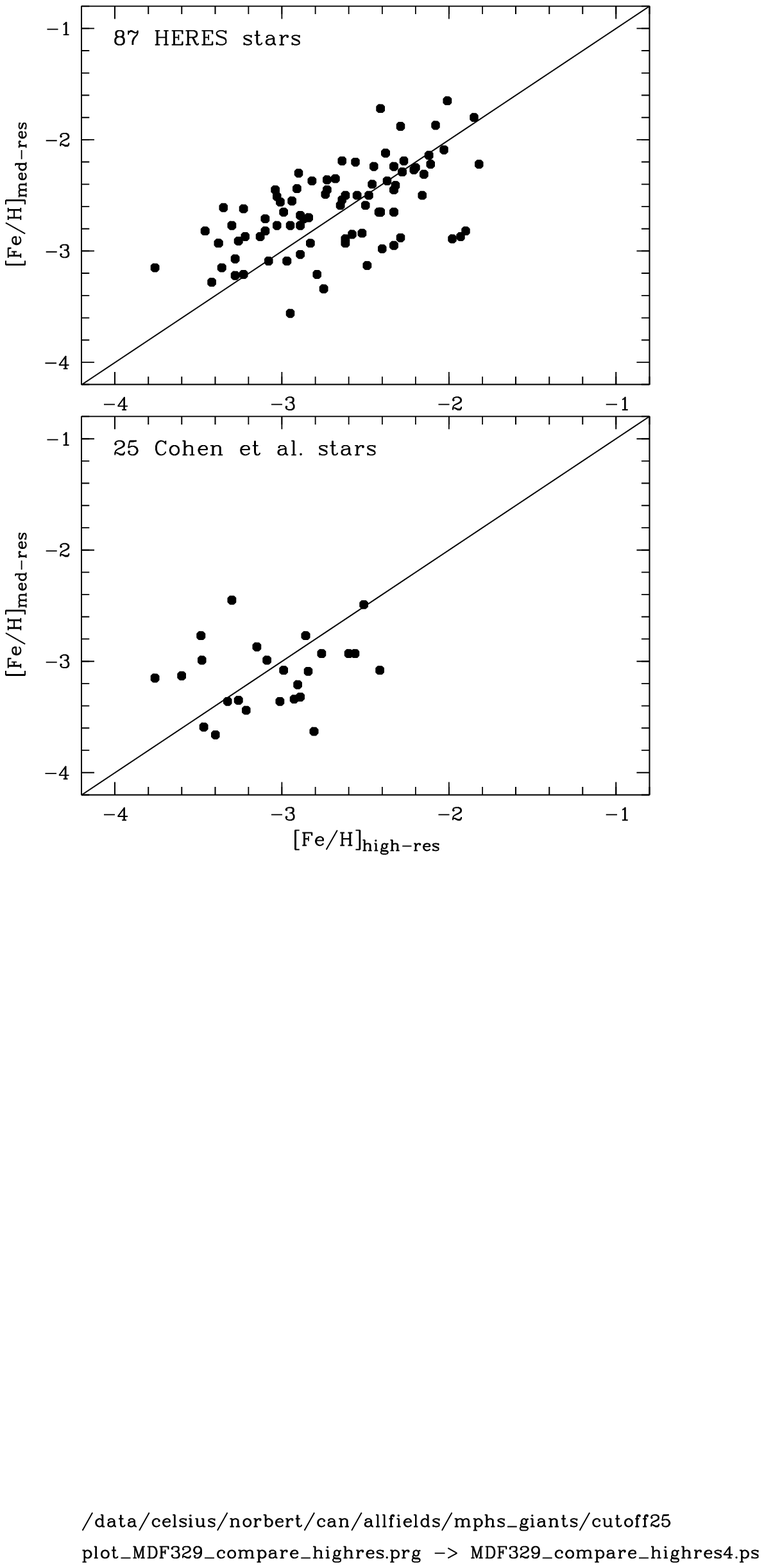}
  \caption{\label{Fig:HighResComparison} Comparison of determinations of
    [Fe/H] from moderate-resolution follow-up spectra with results based on
    high-resolution spectroscopy. Upper panel: 87 stars observed with VLT/UVES 
    \citep{HERESpaperII}. Lower panel: 23 stars observed with Keck/HIRES and
    two with Magellan/MIKE; analyses carried out by \citet{KeckpaperIV},
    \citet{Cohenetal:2006}, \citet{Cohenetal:2008}, and Cohen (2008,
    unpublished).}
\end{figure}

In the follow-up spectra, we measured the KP index as well as the HP2 index of
H$\delta$ and the GP index for the G-band of CH (see \citealt{Beersetal:1999}
for the definition of these indices). When multiple spectra for a star were
available, the $S/N$-weighted average of the individual line index
measurements was adopted. [Fe/H] was determined from the adopted HP2 and KP
indices using the method of \citet{Beersetal:1999}. Since the publication of
that paper, the algorithm was improved mainly by including more calibration
stars, which results in better coverage of the relevant stellar parameter
space, and in particular in a better coverage of the low-[Fe/H] region.

Since the stars of our sample were observed with many telescope/instrument
combinations, it is important to verify that there are no systematic offsets
between the measurements of the line indices in spectra taken at different
telescopes. Such offsets could occur, for example, if the CCD response curves
would strongly vary from instrument to instrument in the wavelength ranges in
which the line and continuum bands of the indices are measured. For this
reason, a number of candidate metal-poor stars were intentionally re-observed
at different telescopes. Furthermore, in most of the observing campaigns,
spectra of a few metal-poor standards (e.g., G~64--12, HD~140283, or {\cd}) as
well as metal-poor radial-velocity standards were secured. In
Fig.~\ref{Fig:KPHP2test}, we show comparisons of the KP and HP2 indices
measured in spectra taken with all relevant telescope/instrument combinations.
In total, 315 pairs of measurements are available. No systematic offsets
between the measurements can be seen. However, the scatter of the measurements
in spectra obtained with the UK~Schmidt and the fibre-fed multi-object
spectrograph 6dF are about a factor two larger than those of the other
telescope/instrument combinations. This can be attributed to the fact that sky
subtraction is more difficult for the 6dF spectra, since only a few fibers
were dedicated to measure the sky background, and furthermore the sky
brightness might have varied over the $6^{\circ}$ diameter field of view of
the instrument.

The quality of the spectra (i.e., $R\sim 2000$ and a typical $S/N$ of 20 per
pixel in the continuum near the Ca~K line) allowed the easy identification and
rejection of emission-line and other ``peculiar'' objects (e.g., galaxies, or
objects with continuous spectra, such as cool, helium-rich white dwarfs). It
has been shown by \citet{Cohenetal:2005} that CH lines present in the
continuum bands of the KP and HP2 indices lead to a systematic underestimation
of these indices, resulting in systematically too low [Fe/H] values. Hence, we
also excluded from this study all stars with $\mathrm{GP} > 6$\,{\AA}. Since
the fraction of carbon-enhanced stars among metal-poor stars increases as the
metallicity decreases (see, e.g.,
\citealt{Cohenetal:2005,Lucatelloetal:2006}), the rejection of stars with
strong G-bands might lead to a bias against low-metallicity stars. However,
since only 90 stars, or 5\,\% of the 1771 observed stars, were rejected due to
this reason, the possible effect on our sample is only minor. We also note
that the three currently-known ultra metal-poor stars (i.e., stars with
$\mathrm{[Fe/H]} < -4.0$; see Sect.~\ref{Sect:MDFconstruction} below), all of
which have large overabundances of carbon, are not rejected by this criterion,
since their GP indices are smaller than 6\,{\AA}. In total, 133 stars were
rejected, leaving 1638.

%
%

Homogeneous abundance analyses based on high-resolution spectra are available
for 112 of the confirmed candidates in our sample. The spectra were taken with
VLT/UVES (87 stars), Keck/HIRES (23 stars) or Magellan/MIKE (2 stars).  The
abundance analyses were performed by \citet{HERESpaperII},
\citet{KeckpaperIV}, \citet{Cohenetal:2006}, \citet{Cohenetal:2008}, and Cohen
(unpublished). Fig.~\ref{Fig:HighResComparison} compares the iron
abundances determined in the course of these analyses
($\mathrm{[Fe/H]}_{\mathrm{high-res}}$) to the moderate-resolution follow-up
results ($\mathrm{[Fe/H]}_{\mathrm{med-res}}$). No significant trends or
offsets are present, and the 1-$\sigma$ scatter around a regression line of
the combined test sample is 0.3\,dex. We hence conclude that the accuracy of
$\mathrm{[Fe/H]}_{\mathrm{med-res}}$ for our sample is $\pm 0.3$\,dex.  We
note that the accuracy can be increased especially for the cooler stars by
using CCD photometry rather than $B-V$ colors predicted from the H$\delta$
index HP2 when deriving $\mathrm{[Fe/H]}_{\mathrm{med-res}}$. However, CCD
photometry is not yet available for all stars of our sample.

To increase the accuracy of the determination of the shape of the
low-metallicity tail of the MDF, we replaced
$\mathrm{[Fe/H]}_{\mathrm{med-res}}$ with
$\mathrm{[Fe/H]}_{\mathrm{high-res}}$, where available.
$\mathrm{[Fe/H]}_{\mathrm{high-res}}$ values are available for 27 of the 76
stars at $\mathrm{[Fe/H]}_{\mathrm{med-res}} < -3.0$, and five out of the six
with $\mathrm{[Fe/H]}_{\mathrm{med-res}} < -3.5$. The
$\mathrm{[Fe/H]}_{\mathrm{high-res}}$ values were taken from the references
above and from \citet{Cayreletal:2004} for HE~0305$-$5442, a re-discovery of
CS~22968-014 ($\mathrm{[Fe/H]}_{\mathrm{high-res}}=-3.56$).  The sixth star at
$\mathrm{[Fe/H]}_{\mathrm{med-res}} < -3.5$ in our sample for which a
$\mathrm{[Fe/H]}_{\mathrm{high-res}}$ estimate is available has
$\mathrm{[Fe/H]}_{\mathrm{med-res}} = -4.2$. A VLT/UVES spectrum exists for
this star, and a preliminary abundance analysis confirms that the star has a
metallicity close to or slightly below $\mathrm{[Fe/H]}=-4.0$. Due to the
preliminary nature of this result, we do not show this data point in
Fig.~\ref{Fig:HighResComparison}.


\begin{figure}[htbp]
  \centering
  \includegraphics[clip=true,bb=105 426 405 628,width=8.5cm]{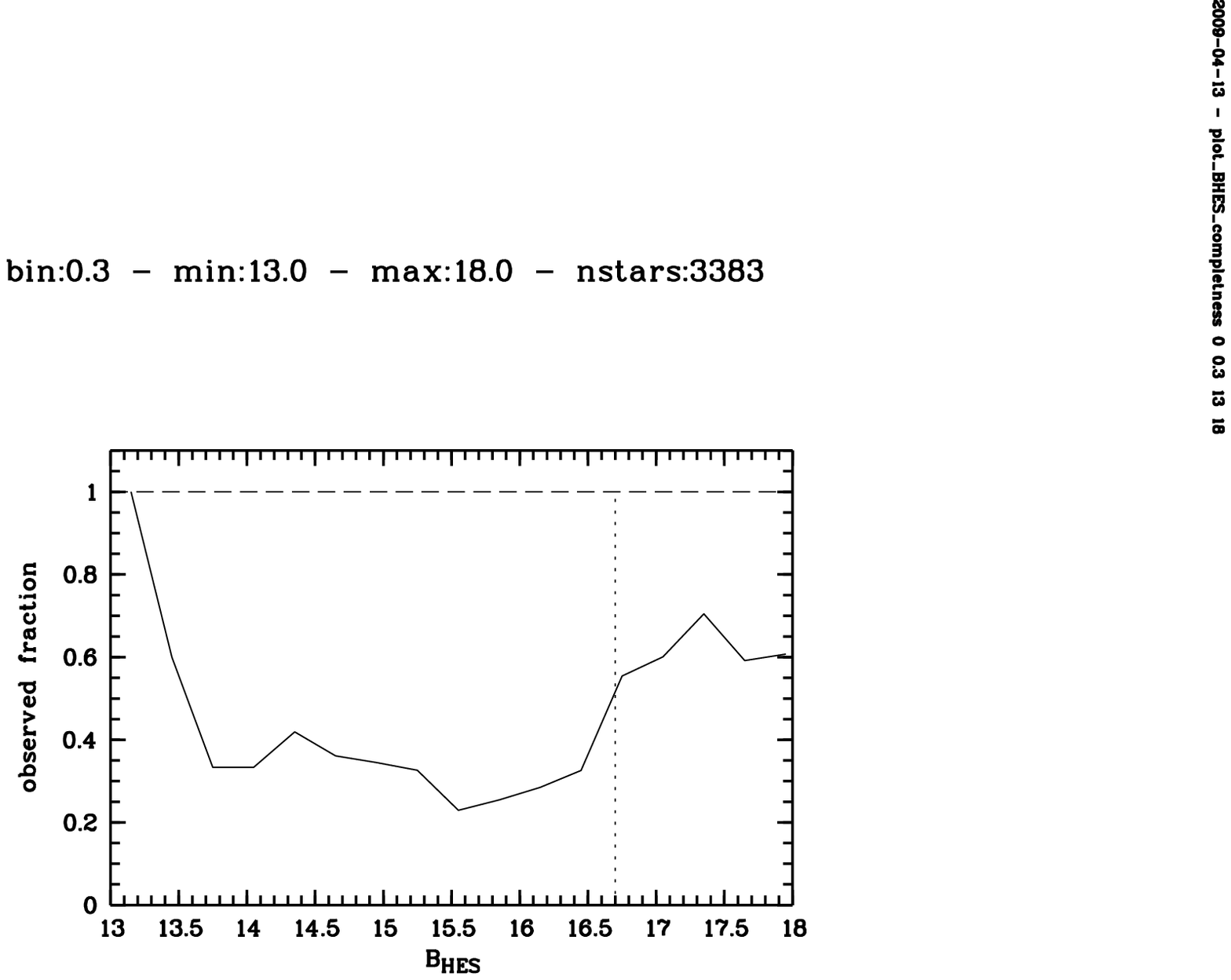}
  \caption{\label{Fig:completeness_Bmag} Fraction of the stars in the HES
    sample for which moderate-resolution follow-up spectroscopy exists as
    function of B magnitude.}
\end{figure}

\section{Construction of the observed MDF}\label{Sect:MDFconstruction}

\begin{figure*}[htbp]
  \centering
  \includegraphics[clip=true,bb=30 341 482 779,width=\textwidth]{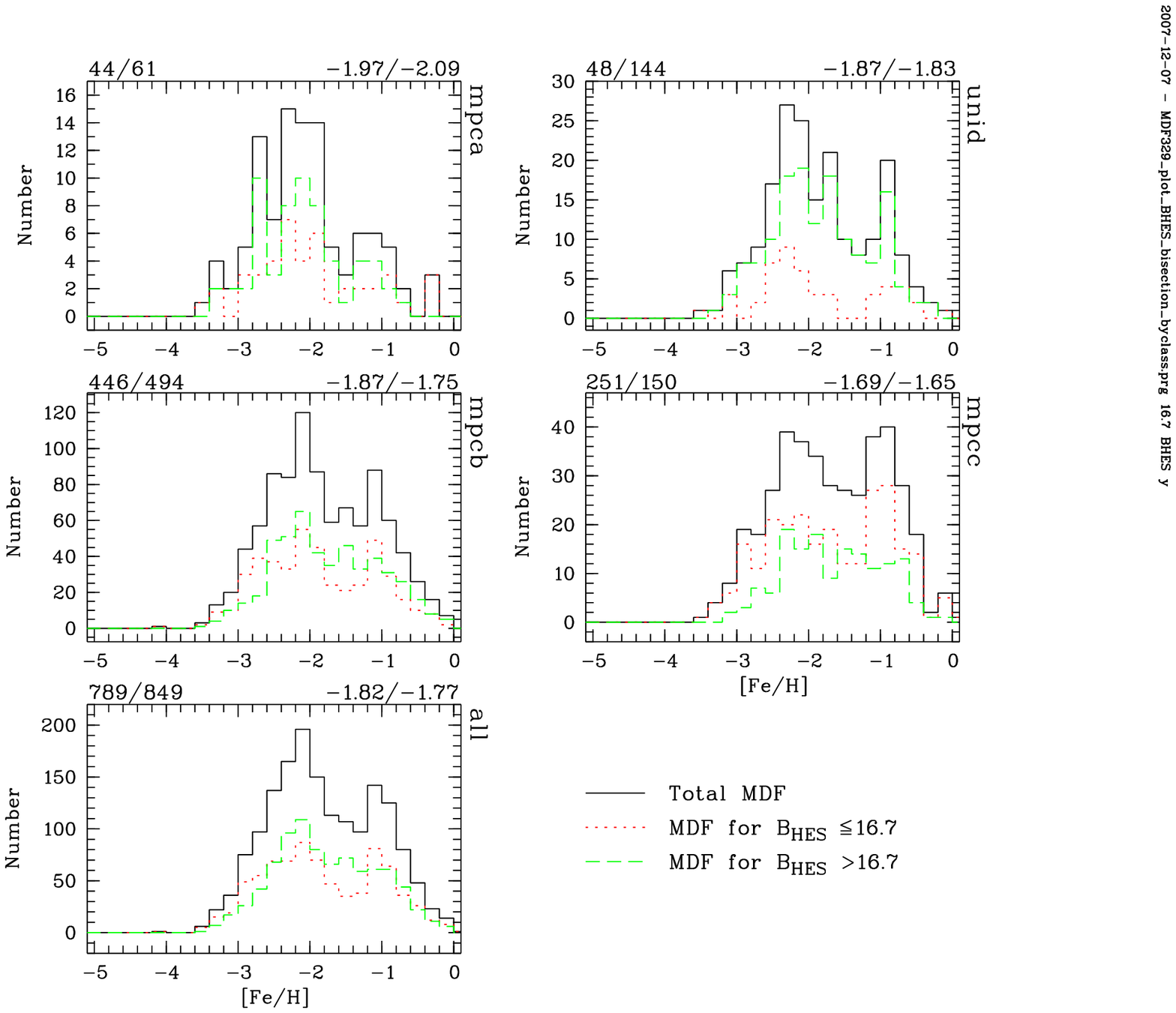}
  \caption{\label{Fig:HES_MDF_Bmag} Comparison of the MDFs of the bright
    ($B \le 16.7$) and faint ($B > 16.7$) subsamples for each of the four
    candidate classes (upper four panels) as well as for the combined
    candidate sample (lower left panel). At the top left of each panel, the
    number of candidates belonging to the bright and faint sample,
    respectively, is listed; at the top right, the mean [Fe/H] of the samples
    is given. }
\end{figure*}

In order to investigate potential selection biases given the adopted follow-up
observation strategy, it is instructive to compare the MDFs derived from stars
of the individual candidate classes and in different magnitude ranges.  For
the purpose of investigating the possible presence of a bias caused by the
fact that the brightest stars were observed first, we divided the full HES
sample, as well as the subsamples of the four candidate classes, into a bright
($B \le 16.7$) and a faint ($B > 16.7$) half, respectively. The results are
shown in Fig.~\ref{Fig:HES_MDF_Bmag}.

In the upper right panel it can be seen that the faint candidates are
over-represented in the class \texttt{unid}. The reason is that the visual
classification for fainter candidates, which have lower quality HES spectra,
was more difficult. The bright- and faint-star [Fe/H] distributions of the
other candidate classes, as well as the total sample, appear very similar to
one another. This is quantitatively confirmed for the \texttt{mpca} and
\texttt{unid} subsets by means of a Kolmogorov-Smirnov (KS) test of the null
hypothesis $H_0$ that the bright and faint subsets of the stars belonging to
these candidate classes were drawn from the same parent distribution, against
the alternative hypothesis $H_1$ that they were not drawn from the same parent
distribution. We chose a significance level of $\alpha=0.01$; i.e., $H_0$ is
rejected if the probability $p$ of occurence of the test statistic (i.e., in
case of the KS-test, D, the maximum distance between the cumulative
probability distributions of the two samples), given $H_0$, is smaller than
0.01. The result of the KS-test for the bright and faint stars of the classes
\texttt{mpca} and \texttt{unid} are $p=0.82$ and $p=0.21$, respectively; i.e.,
$H_0$ can clearly not be rejected in these cases.  However, for the other two
classes, the probabilities are considerably lower, ranging from 0.0046
(\texttt{mpcb}) to 0.073 (\texttt{mpcc}). For the full sample (i.e., all
candidate classes combined), the probability is 0.0028; that is, the
distributions of the bright and faint subsamples differ significantly from
each other. It would hence be desirable to construct the halo MDF from the
bright and faint samples separately. However, in this case the sample sizes
would be too small to draw any conclusions, in particular about the
low-metallicity tail of the MDF. Therefore, we decided to construct the halo
MDF from the full sample. We note that the relative fraction of observed stars
does not exhibit any strong biases towards bright or faint stars (see
Fig.~\ref{Fig:completeness_Bmag}), thus the sample from which we construct the
MDF should at least be representative for halo stars in the magnitude range
$13 < B < 17.5$.

\begin{table*}[htbp]
 \centering
 \caption{\label{Tab:observedMDF} The MDF of the Galactic halo field stars as 
   constructed from the sample of 1638 HES with available spectroscopic 
   follow-up observations, by means of scaling to the full candidate sample of 
   3439 stars (for details see text). Note that for a proper comparison with 
   the MDFs predicted by theoretical models, or the MDFs of other stellar 
   populations, the selection efficiency of the HES as a function of [Fe/H] and 
   $(B-V)_0$ must be taken into account (see Tab.~\ref{Tab:selfrac}). Note
   also that the HES sample is increasingly contaminated with thin- and 
   thick-disk stars as [Fe/H] approaches $0.0$.
}

  \begin{tabular}{r|rrrrrrrrrrrrr}\hline\hline
    \rule{0ex}{2.3ex} [Fe/H] & $-4.50$ & $-4.30$ & $-4.10$ & $-3.90$ & $-3.70$ 
                             & $-3.50$ & $-3.30$ & $-3.10$ & $-2.90$ & $-2.70$ 
                             & $-2.50$ & $-2.30$ & $-2.10$\\
                         $N$ &      0  &      0  &      2  &      0  &      0  
                             &     12  &     45  &     73  &    160  &    198  
                             &    281  &    337  &    399\\\hline
    \rule{0ex}{2.3ex} [Fe/H] & $-1.90$ & $-1.70$ & $-1.50$ & $-1.30$ & $-1.10$ 
                             & $-0.90$ & $-0.70$ & $-0.50$ & $-0.30$ & $-0.10$ 
                             & $+0.10$ & $+0.30$ & $+0.50$\\
                         $N$ &    313  &    231  &    229  &    209  &    308  
                             &    268  &    178  &    109  &     45  &     33  
                             &      3  &      6  &      0 \\\hline
  \end{tabular}
\end{table*}

\begin{figure}[htbp]
  \centering
  \includegraphics[clip=true,bb=77 128 353 770,width=8.5cm]{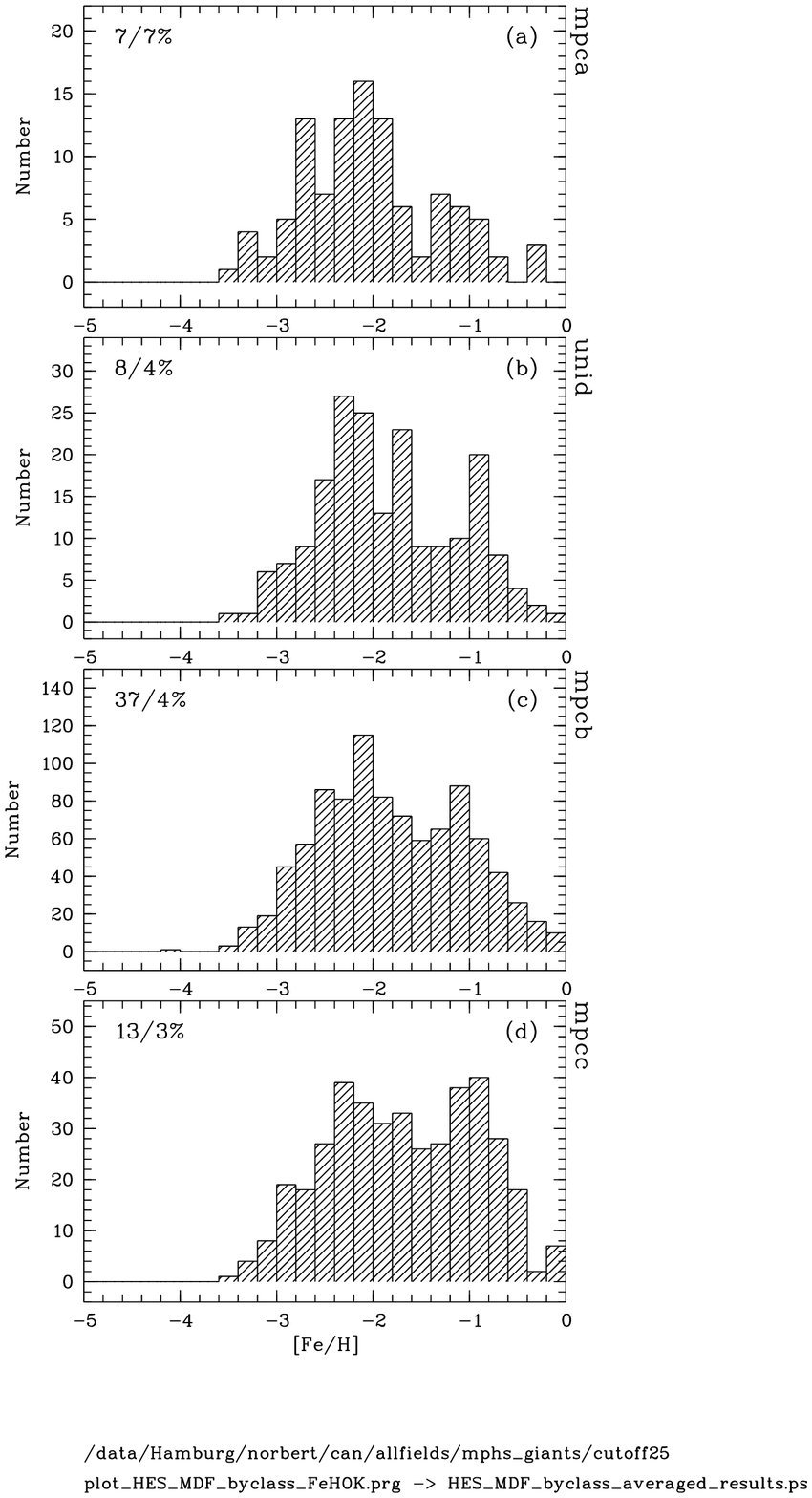}
  \caption{\label{Fig:HES_MDF_byclass} Metallicity distribution of the HES
    sample of 1638 stars, divided by candidate class. In the upper left
    corner of each panel, the number of stars with $\mbox{[Fe/H]}<-3.0$ and
    the percentage of such stars within each candidate class is indicated. }
\end{figure}

As can be seen in Fig.~\ref{Fig:HES_MDF_byclass}, the fraction of stars at
$\mbox{[Fe/H]}<-3.0$ is highest among the \texttt{mpca} candidates (i.e.,
7\,\%), and significantly lower in the other classes (i.e., 3--4\,\%).  That
is, the MDF of \texttt{mpca} candidates is biased towards lower metallicities.
KS-tests show that the null hypotheses $H_0$ that subsamples of different
candidate classes were drawn from the same parent distribution can clearly not
be rejected for neighbouring classes (e.g., $p=0.34$ for \texttt{unid} versus
\texttt{mpcb}), while $H_0$ can be rejected at high significance when more
distant classes are compared to each other (e.g., $p=1.7\cdot 10^{-5}$ for
\texttt{mpca} versus \texttt{mpcc}). These tests and the bias of the
candidates of class \texttt{mpca} towards low metallicity demonstrate that the
candidate classification effectively separated the ``good'' from ``bad''
candidates.  Fig.~\ref{Fig:HES_MDF_byclass} also shows that the number of
false positives (i.e., stars at $\mbox{[Fe/H]}>-2.5$) is considerably higher
among the \texttt{mpcc} candidates. However, this contamination does not
affect our study, because we are mainly concerned with the low-metallicity
tail of the MDF.

In order to properly take into account the stars of our candidate sample for
which no spectroscopic follow-up observations exist, we constructed MDFs from
the observed sample of stars in the following two ways. First, we computed
separate MDFs for each of the candidate classes and scaled them such that the
correct relative fraction of stars is produced when the four scaled MDFs are
coadded; i.e., the scaling factors listed in the last column of
Tab.~\ref{Tab:CandidateStatistics} were applied. Secondly, we assigned to each
of the 1942 stars in the full candidate sample lacking follow-up observations
the [Fe/H] of a randomly selected star of the same candidate class for which a
follow-up spectrum \emph{is} available. We also randomly rejected stars with a
too strong G-band and ``peculiar'' stars according to the probabilities
determined from the sample for which follow-up observations exist. In this
way, a sample of 3439 stars with the correct relative fraction of the
candidates of the four classes was created.

\begin{figure}[htbp]
  \centering
  \includegraphics[clip=true,bb=75 448 471 713,width=8.8cm]{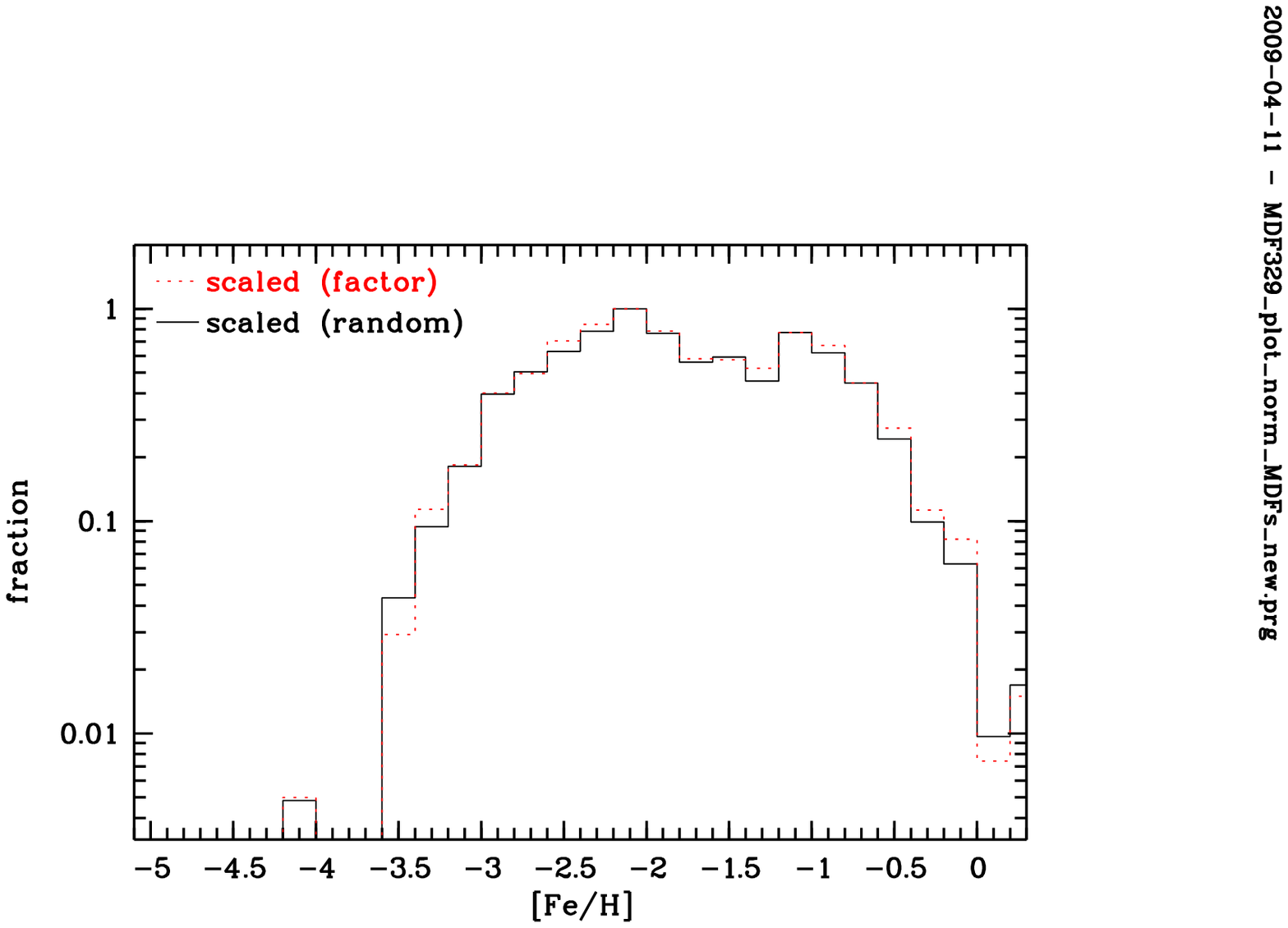}
  \caption{\label{Fig:RandomvsFactor} Comparison of the MDFs
      constructed from the HES sample by means of random scaling and
      co-addition of the class-wise MDFs (solid black line) and scaling by
      factors (grey dotted line).}
\end{figure}

The MDFs produced by these two methods are expected to be very similar to each
other, because in each of them, the class-wise MDFs are scaled and then added
to produce the final MDF; only the scaling methods are slightly different.
Indeed, as can be seen in Fig.~\ref{Fig:RandomvsFactor}, the results do not
differ significantly from each other. A $\chi^2$-test of the null hypotheses
$H_0$ that the two samples are drawn from the same distribution yields a
probability of $p=1.0$; i.e., $H_0$ can very clearly not be rejected.  We
adopt the MDF constructed by means of scaling the class-wise MDFs by a factor
and adding them up. For this MDF, the numbers of stars in each metallicity bin
are listed in Tab.~\ref{Tab:observedMDF}.

Note that we have not corrected the MDF for the fact that as metallicity
decreases, given that the giants become brighter and the dwarfs fainter (see
Fig.~\ref{Fig:VmagBminV}), the relative volumes surveyed in our magnitude
limited sample become larger and smaller, respectively. At $(B-V)_0 = 0.6$,
for example, the data of the Yale-Yonsei isochrones \citep{Kimetal:2002} for
an age of 12\,Gyr suggest that the ratios of volume surveyed at
$\mathrm{[Fe/H]}=-3$ to that at $\mathrm{[Fe/H]}=-2$ are 3.0 and 0.67 for
giants and dwarfs, respectively. Due to the very small survey volume for
dwarfs, no cool main-sequence star has so far been identified in the HES, and
the sample considered in this paper is dominated by giants.  Therefore, a
survey volume correction would lead to a reduced relative number of stars at
the lowest metallicities.

\begin{figure}[htbp]
  \centering
  \includegraphics[clip=true,bb=75 448 471 713,width=8.8cm]{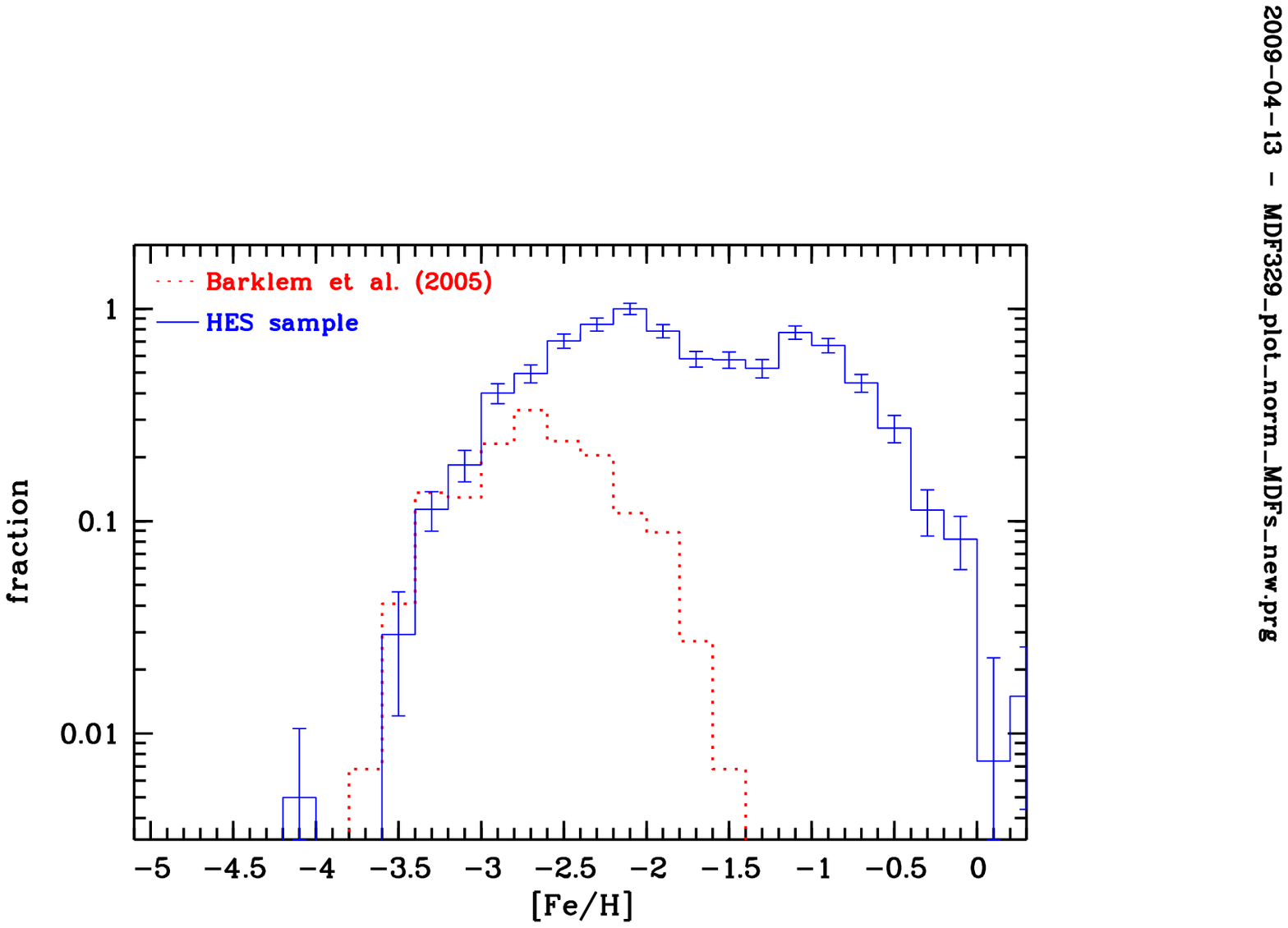}
  \caption{\label{Fig:BarklemvsHES} Comparison of the MDF constructed
      from the HES sample (solid line) with that of the HERES sample analysed by
    \citet[][grey dotted line]{HERESpaperII}. The latter sample is biased
    against stars at $\mathrm{[Fe/H]}>-3.0$, because most stars at higher
    metallicity were intentionally removed due to the science aims of that
    survey.}
\end{figure}

\section{The shape of the halo MDF}\label{Sect:MDFshape}

A prominent feature in both of the scaled MDFs is a sharp drop at
$\mbox{[Fe/H]}\sim -3.6$ (see Fig.~\ref{Fig:RandomvsFactor}); in our (scaled)
sample, only two out of 3439 stars have $\mbox{[Fe/H]}< -3.6$. Such a drop was
also recognized by \cite{JohnTSS}, and it has been seen in the Hamburg/ESO
R-process Enhanced star Survey (HERES; see Fig.~2 of \citealt{HERESpaperII}
and our Fig.~\ref{Fig:BarklemvsHES}). It reflects the fact that only very few
stars at $\mbox{[Fe/H]}< -3.6$ were found in projects aiming at the
identification and detailed study of the lowest metallicity stars of the
Galactic halo, despite the considerable effort expended to find them (see,
e.g., \citealt{Cohenetal:2008}, and references therein).

\begin{figure}[htbp]
  \centering
  \includegraphics[clip=true,bb=85 419 486 685,width=8.8cm]{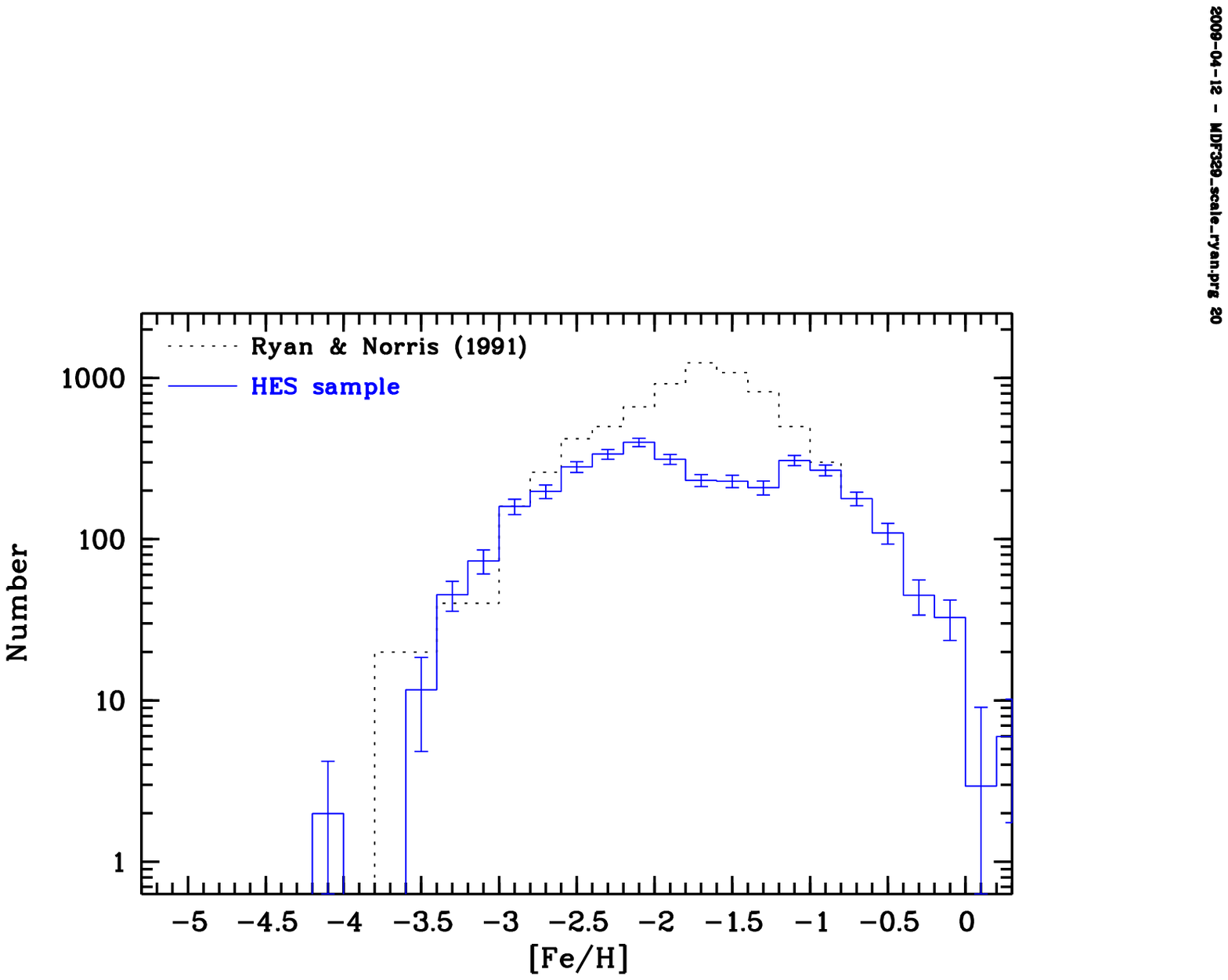}
  \caption{\label{Fig:HESvsRN} Comparison of the halo MDF constructed from the
    HES sample (histogram) with that of \citet{Ryan/Norris:1991}, scaled to
     match the HES MDF in the range $-3.4 < \mathrm{[Fe/H]} < -2.5$ (dotted line).}
\end{figure}

The shape of the low-metallicity end of the halo MDF could not be determined
precisely by \citet{Ryan/Norris:1991} due the limited size of their sample,
which contains only four stars at $\mathrm{[Fe/H]} < -3.4$, and none with
$\mathrm{[Fe/H]} < -4.0$. As can be seen in Fig.~\ref{Fig:HESvsRN}, in the
range $-3.4 < \mathrm{[Fe/H]} < -2.5$ their halo MDF agrees very well with the
HES MDF. In Fig.~\ref{Fig:HESvsRN} one can see a disagreement between the two
MDFs in the bin centered on $\mathrm{[Fe/H]} = -3.5$; i.e., the number of
stars at this metallicity in the sample of Ryan \& Norris is higher by about a
factor of two as compared to the HES sample. Alternatively, the number of
stars in the range $-3.4 < \mathrm{[Fe/H]} < -2.5$ (i.e., the metallicity
range which has been used to scale the two MDFs onto each other) are
under-represented in the sample of Ryan \& Norris, or over-represented in the
HES sample. Even though the number of stars at $\mathrm{[Fe/H]} < -3.4$ in
both samples is small, the difference is significant. A KS-test of the null
hypothesis $H_0$ that the HES sample and the sample of Ryan \& Norris have the
same parent distribution at $\mathrm{[Fe/H]} < -2.5$ yields a probability of
$p=0.0087$; i.e., $H_0$ must (barely) be rejected if $\alpha = 0.01$ is
chosen. The probability increases to 0.0091 if the two stars at
$\mathrm{[Fe/H]} < -4.0$ are excluded from the HES sample. The reason for the
discrepancy is currently unclear, but one might speculate that it is related
to the kinematic selection of the sample of Ryan \& Norris and/or a larger
fraction of stars belonging to the outer halo population being present in the
HES sample.

Another feature of the halo MDF is a lightly populated tail extending to
$\mbox{[Fe/H]}< -5.0$. The evidence for this feature from our (scaled) sample
alone is weak, since it contains only two stars at $\mbox{[Fe/H]}< -3.6$, and
none at $\mbox{[Fe/H]}< -4.3$. However, currently some ten stars with
$\mbox{[Fe/H]}< -3.6$ have published abundance analyses based on
high-resolution spectroscopy (see Tab.~4 of \citealt{Beers/Christlieb:2005}
for a recent review), including three additional stars at $\mbox{[Fe/H]}<
-4.0$: {\FAC} ($\mathrm{[Fe/H]} = -5.4$;
\citealt{Frebeletal:2005,Aokietal:2006,Frebeletal:2006a}), {\CBB}
($\mathrm{[Fe/H]} = -5.3$;
\citealt{HE0107_Nature,HE0107_ApJ,Besselletal:2004}), and HE~0557$-$4840
($\mathrm{[Fe/H]} = -4.8$; \citealt{Norrisetal:2007}). These three stars are
not part of our sample due to a variety of reasons. {\FAC} is part of the
bright HES metal-poor sample consisting of stars above a saturation threshold
\citep{Frebeletal:2006b}, while only unsaturated point sources entered the
sample of this work. {\CBB} was selected in a previous version of the
candidate selection which was slightly less restrictive than the one we use
here; as a result, this star misses the selection cutoff of 3.9\,{\AA} for its
HES $(B-V)_0$ colour of $0.6$\,mag by 0.1\,{\AA} (i.e., the KP index measured
in the HES spectrum is 4.0\,{\AA}). And finally, HE~0557$-$4840 is located on
one of the 50 HES plates which are not considered here. In conclusion, for an
accurate determination of the shape of the MDF at $\mathrm{[Fe/H]} < -4.0$ it
is required to compile even larger statistically complete samples of
metal-poor stars.

%


\section{Comparison between theoretical MDFs and the halo MDF}\label{Sect:TheoryObservations}

In a comparison of the observed MDF with MDFs predicted by theoretical models,
one has to take into account the modification of the shape of the MDF by the
selection of metal-poor candidates employed in the HES. In particular,
uncertainties $\sigma_{\mathrm{KP}}$ and $\sigma_{B-V}$ of the measurements of
the KP index and $B-V$ in the HES spectra result in a scatter of stars with
$\mbox{[Fe/H]} > -2.5$ into the sample, and stars with $\mbox{[Fe/H]} < -2.5$
out of the sample. Each theoretical MDF under investigation is therefore
converted into an MDF as it would be observed in the HES, by applying the
metal-poor star selection criteria used in the HES.

\begin{figure}[htbp]
  \centering
  \includegraphics[clip=true,bb=98 426 400 628,width=8.8cm]{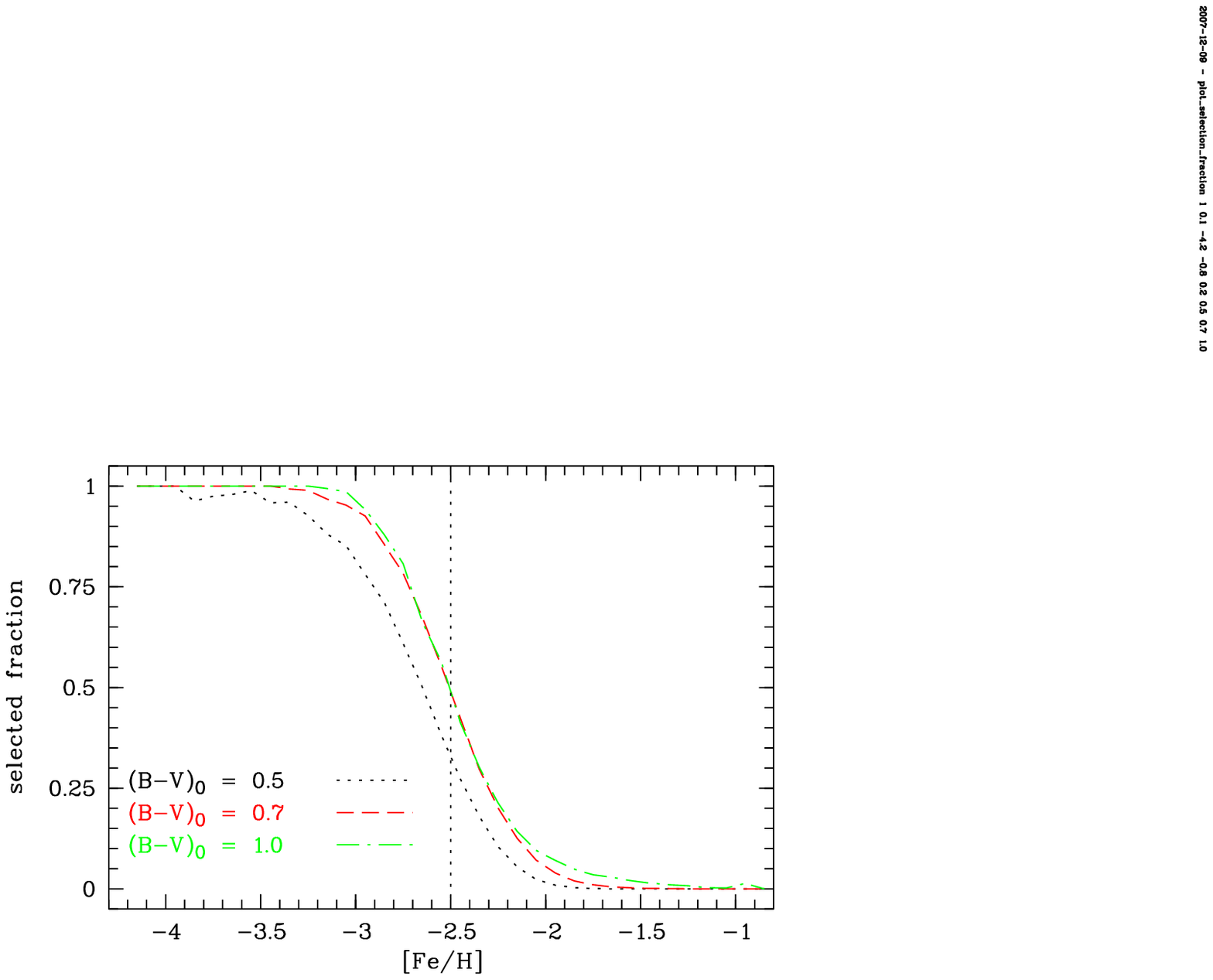}
  \caption{\label{Fig:selfrac} Selection function for HES metal-poor 
    candidates of $(B-V)_0=0.5$, $0.7$, and $1.0$, as determined from
      a simulated sample of stars following the MDF predicted by the Simple
      Model of Galactic chemical evolution \citep{Searle/Sargent:1972,Hartwick:1976}.}
\end{figure}

\begin{table}[htbp]
 \centering
 \caption{\label{Tab:selfrac} Selection function for HES metal-poor candidates
   in the colour range $0.5 < (B-V)_0 < 1.0$, as determined from a simulated 
   sample of stars following the MDF predicted by the Simple Model of Galactic 
   chemical evolution \citep{Searle/Sargent:1972,Hartwick:1976}.
}
  \begin{tabular}{lcccccc}\hline\hline
                & \multicolumn{6}{c}{Selected fraction at $(B-V)_0$}\\
    \rb{[Fe/H]} & 0.5   & 0.6   & 0.7   & 0.8   & 0.9   & 1.0 \\\hline
        $-4.05$ & 1.000 & 1.000 & 1.000 & 1.000 & 1.000 & 1.000\rule{0.000ex}{2.3ex}\\
        $-3.95$ & 0.958 & 1.000 & 1.000 & 1.000 & 1.000 & 1.000\\
        $-3.85$ & 1.000 & 0.962 & 1.000 & 1.000 & 1.000 & 1.000\\
        $-3.75$ & 0.961 & 1.000 & 1.000 & 1.000 & 1.000 & 1.000\\
        $-3.65$ & 0.982 & 0.987 & 1.000 & 1.000 & 1.000 & 1.000\\
        $-3.55$ & 0.954 & 0.991 & 1.000 & 1.000 & 1.000 & 1.000\\
        $-3.45$ & 0.920 & 0.991 & 0.997 & 0.996 & 1.000 & 1.000\\
        $-3.35$ & 0.924 & 0.979 & 0.993 & 0.996 & 0.997 & 1.000\\
        $-3.25$ & 0.901 & 0.976 & 0.989 & 0.991 & 0.985 & 1.000\\
        $-3.15$ & 0.861 & 0.950 & 0.984 & 0.974 & 0.981 & 0.997\\
        $-3.05$ & 0.816 & 0.919 & 0.958 & 0.954 & 0.953 & 0.983\\
        $-2.95$ & 0.744 & 0.869 & 0.928 & 0.908 & 0.900 & 0.949\\
        $-2.85$ & 0.668 & 0.801 & 0.879 & 0.852 & 0.839 & 0.918\\
        $-2.75$ & 0.563 & 0.700 & 0.812 & 0.768 & 0.743 & 0.822\\
        $-2.65$ & 0.455 & 0.583 & 0.715 & 0.658 & 0.617 & 0.709\\
        $-2.55$ & 0.340 & 0.457 & 0.592 & 0.537 & 0.488 & 0.573\\
        $-2.45$ & 0.232 & 0.337 & 0.462 & 0.406 & 0.364 & 0.433\\
        $-2.35$ & 0.140 & 0.234 & 0.331 & 0.297 & 0.264 & 0.318\\
        $-2.25$ & 0.075 & 0.149 & 0.222 & 0.203 & 0.187 & 0.217\\
        $-2.15$ & 0.034 & 0.088 & 0.136 & 0.130 & 0.127 & 0.152\\
        $-2.05$ & 0.013 & 0.046 & 0.079 & 0.081 & 0.085 & 0.100\\
        $-1.95$ & 0.004 & 0.021 & 0.043 & 0.048 & 0.058 & 0.070\\
        $-1.85$ & 0.001 & 0.009 & 0.022 & 0.028 & 0.041 & 0.051\\
        $-1.75$ & 0.000 & 0.003 & 0.011 & 0.017 & 0.028 & 0.036\\
        $-1.65$ & 0.000 & 0.001 & 0.006 & 0.010 & 0.020 & 0.028\\
        $-1.55$ & 0.000 & 0.000 & 0.003 & 0.006 & 0.015 & 0.021\\
        $-1.45$ & 0.000 & 0.000 & 0.001 & 0.004 & 0.010 & 0.017\\
        $-1.35$ & 0.000 & 0.000 & 0.001 & 0.003 & 0.007 & 0.014\\
        $-1.25$ & 0.000 & 0.000 & 0.001 & 0.002 & 0.004 & 0.008\\
        $-1.15$ & 0.000 & 0.000 & 0.000 & 0.001 & 0.002 & 0.008\\
        $-1.05$ & 0.000 & 0.000 & 0.000 & 0.001 & 0.001 & 0.004\\
        $-0.95$ & 0.000 & 0.000 & 0.000 & 0.000 & 0.000 & 0.000\\
        $-0.85$ & 0.000 & 0.000 & 0.000 & 0.000 & 0.000 & 0.000\\\hline
  \end{tabular}
\end{table}

The first step in the conversion of a theoretical MDF is the simulation of a
sample of stars with a distribution in [Fe/H] according to that of the
theoretical MDF under investigation. The [Fe/H] values are then converted into
pairs of KP and $(B-V)_0$ by inverting the calibrations of
\citet{Beersetal:1999}. Then, a subsample was selected such that it follows
the distribution in $(B-V)_0$ of the HES sample (see
Fig.~\ref{Fig:VmagBminV}). Taking into account the distribution in $(B-V)_0$
is important because the shape of the selection function is determined by
$\sigma_{\mathrm{KP}}$, $\sigma_{B-V}$, and the gradient of [Fe/H] in the KP
versus $(B-V)_0$ parameter space (see Fig.~4 of Paper~IV); it varies with
$(B-V)_0$, as can be seen in Fig.~\ref{Fig:selfrac}.

The reader will note that we have excluded stars with $(B-V)_0 < 0.5$ from our
sample, which in principle will affect the relative proportions of dwarfs
admitted to our sample as a function of [Fe/H]. In practice, however, this is
not a serious effect if we restrict our abundance range to abundances
$\mathrm{[Fe/H]} < -2.0$.  Consideration of the Yale-Yonsei isochrones for an
age of 12\,Gyr, and for the Salpeter mass function ($x = 1.35$), shows that
the percentages of dwarfs with $(B-V)_0 < 0.5$ relative to all main-sequence
stars with mass greater that 0.40\,M$_{\odot}$ are 4, 13, 19, 22, 24, and 24
for $\mathrm{[Fe/H]} = -1.0$, $-1,5$, $-2.0$, $-2.5$, $-3.0$, and $-3.5$,
respectively.  That is to say, the proportion of excluded dwarfs is relatively
constant for $\mathrm{[Fe/H]} < -2.0$.

The next step in the procedure of converting theoretical MDFs into an MDF as
it would be observed in the HES was to add random Gaussian errors with
standard deviations according to the known measurement uncertainties
$\sigma_{\mathrm{KP}}$, $\sigma_{B-V}$ in the HES to KP and $(B-V)_0$ assigned
to each star. Finally, the KP/$(B-V)_0$ selection criterion was applied to the
simulated sample of stars. The [Fe/H] distribution of the selected stars is
the MDF as it would be observed in the HES.

For the convenience of the reader, we list in Tab.~\ref{Tab:selfrac} and show
in Fig.~\ref{Fig:selfrac} the HES metal-poor star selection function as
determined with a simulated sample of stars following the MDF predicted by the
Simple Model of Galactic chemical evolution
\citep{Searle/Sargent:1972,Hartwick:1976}. That model assumes that a fiducial
``closed box'' of primordial gas is enriched by successive stellar
generations. Further model assumptions are that (i) the gas is well-mixed at
all times (i.e there is a unique age-metallicity relation for the stars formed
from that gas) and (ii) the stellar initial mass function (IMF) does not
change with time. Analytical solutions can only be obtained if it is assumed
that the evolutionary timescales of the enriching stars are neglible (the
so-called Instantaneous Recycling Approximation or IRA). Such solutions can be
generically obtained in the case of a closed box, and in some particular cases
of outflow (gas loss from the box) and infall (gas flows into the box). Since
the IRA turns out to be a very good approximation for elements ejected by
massive stars, those analytical solutions can provide a powerful tool for the
study of Galactic systems.

In the framework of the Simple Model, the shape of the MDF can be described in
terms of a unique parameter, the ``yield'', which is the ratio of the mass of
newly-created metals to the mass locked in long-lived stars and stellar
remnants. This is a very useful parametrization, because it is independent of
the star formation history of the system (the major unknown in Galactic
evolution studies). In the closed box model the yield depends only on the IMF
(referred to as the ``true yield''), while in the case of gaseous flows
(infall and outflow) it depends also on their magnitude; this ``effective
yield'', $y_{\mbox{\scriptsize eff}}$, is always smaller than the true yield.
It turns out that the MDF peaks at a metallicity equal to the effective yield;
this simple result allows one to determine the effective yield and to
constrain the underlying physics (IMF, outflow rate, etc.)

\begin{figure}[htbp]
  \centering
  \includegraphics[clip=true,bb=85 420 485 685,width=8.8cm]{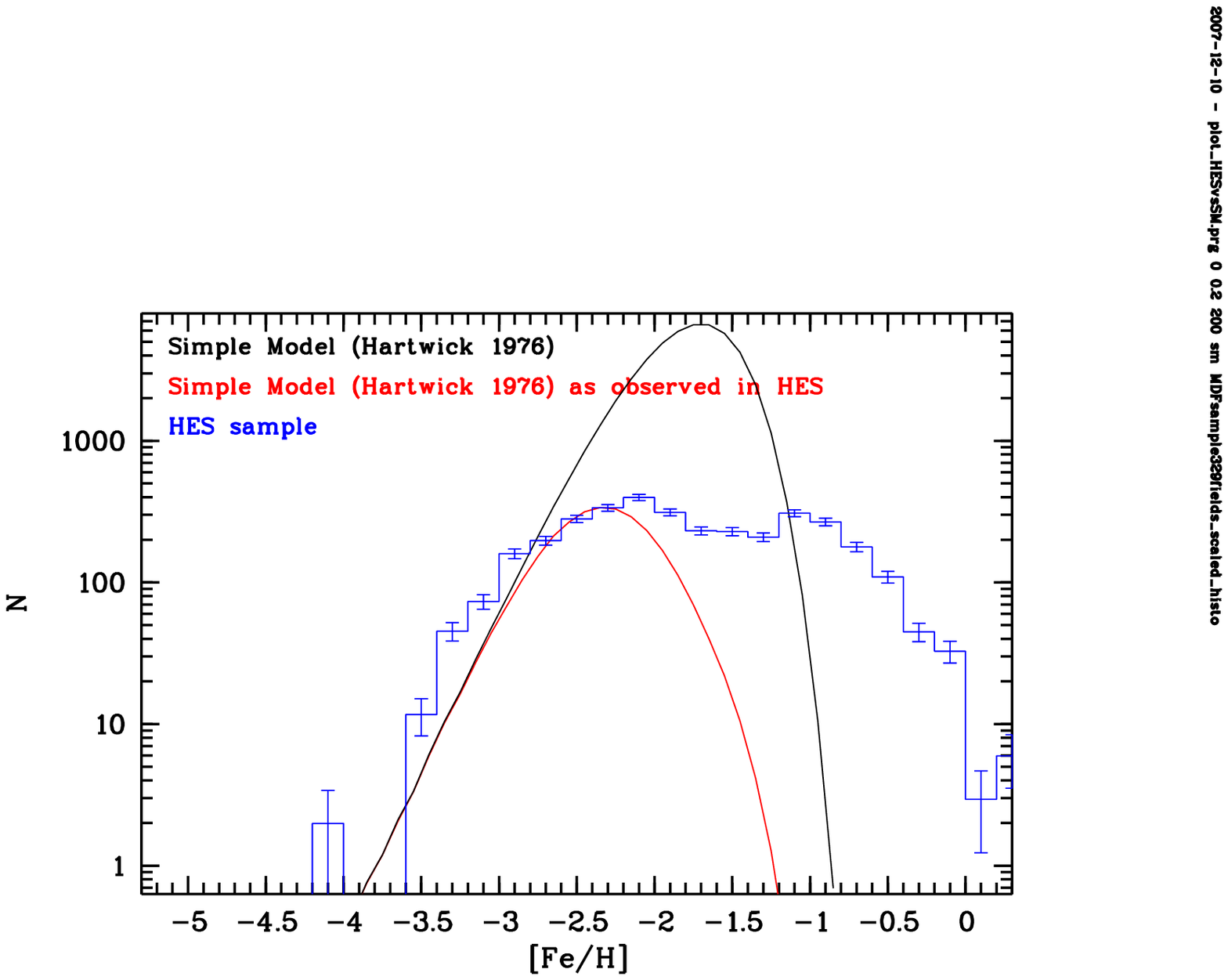}
  \caption{\label{Fig:MDF_SimpleModel} Comparison of the MDF of a Simple Model
    with $y_{\mbox{\scriptsize eff}} = -1.7$ (black line) with the MDF
    constructed from the HES sample (histogram). The grey line illustrates how
    the MDF of the Simple Model would be observed in the HES. For the
    comparison between the theoretical model and the observations, only the
    grey line in the metallicity range below $\mbox{[Fe/H]}\sim -2.0$ should
    be taken into account, because at higher [Fe/H] the HES sample is
    contaminated with thin- and thick-disk stars.}
\end{figure}

\begin{figure}[htbp]
  \centering
  \includegraphics[clip=true,bb=85 420 485 685,width=8.8cm]{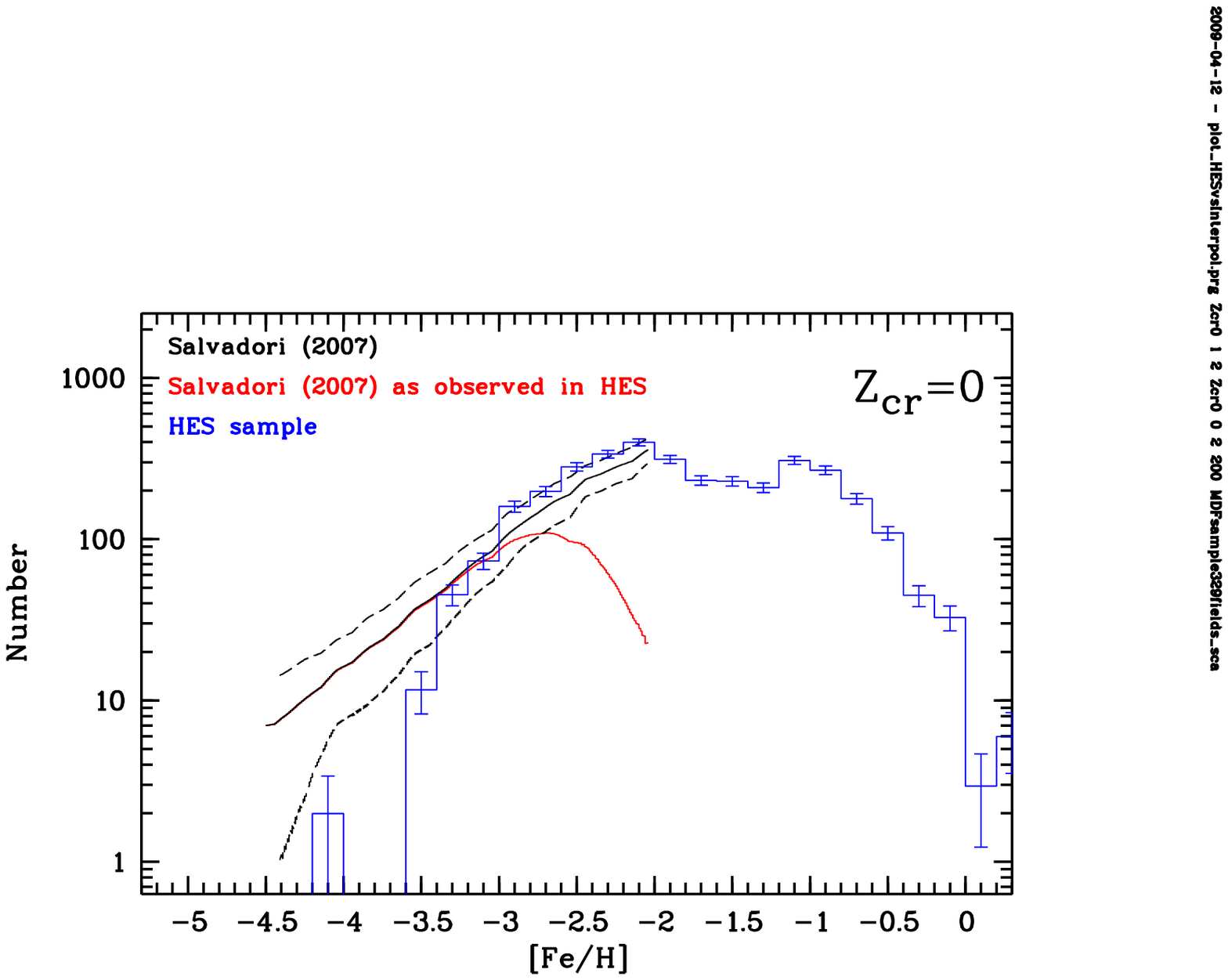}\\
  \includegraphics[clip=true,bb=85 420 485 685,width=8.8cm]{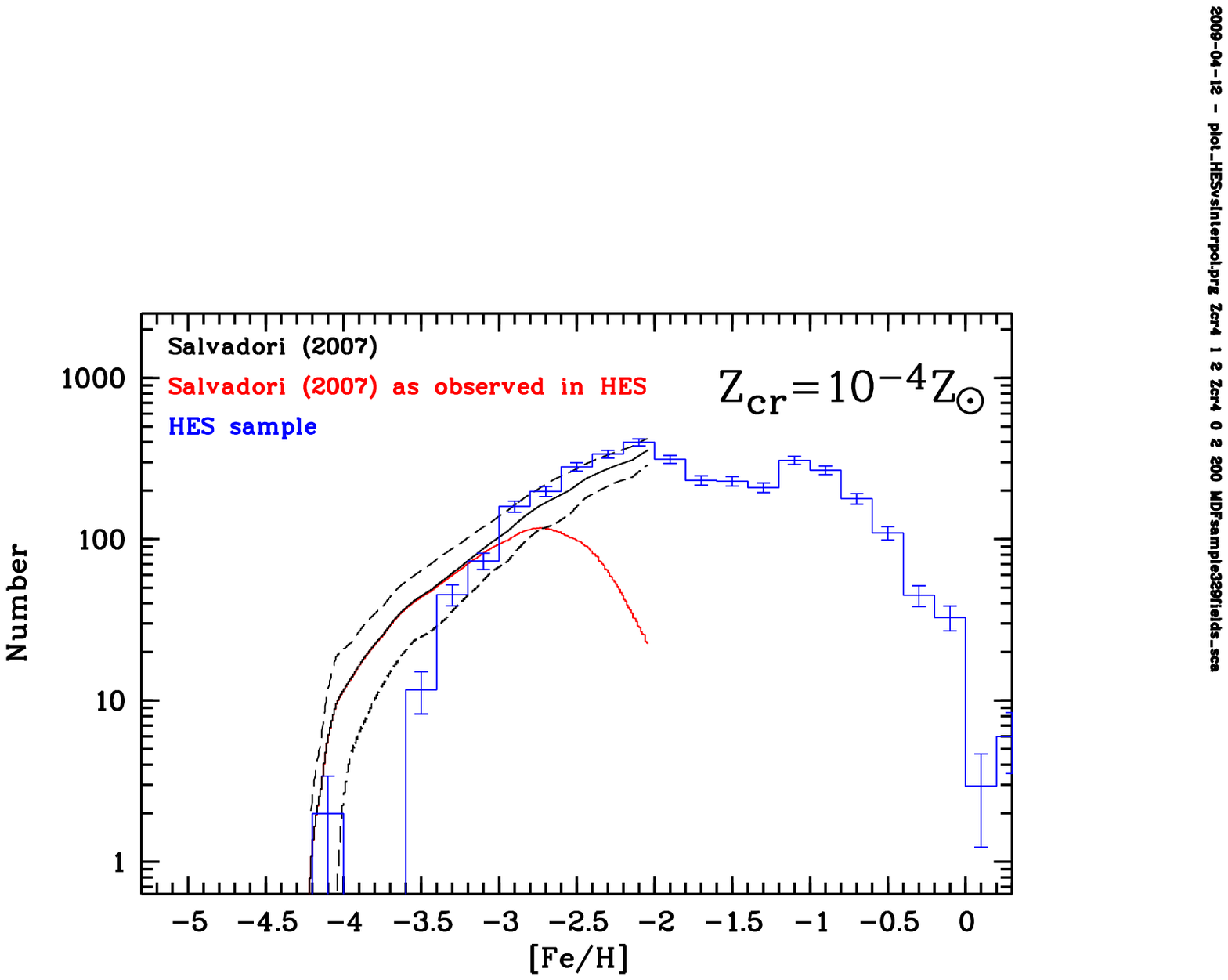}\\
  \includegraphics[clip=true,bb=85 420 485 685,width=8.8cm]{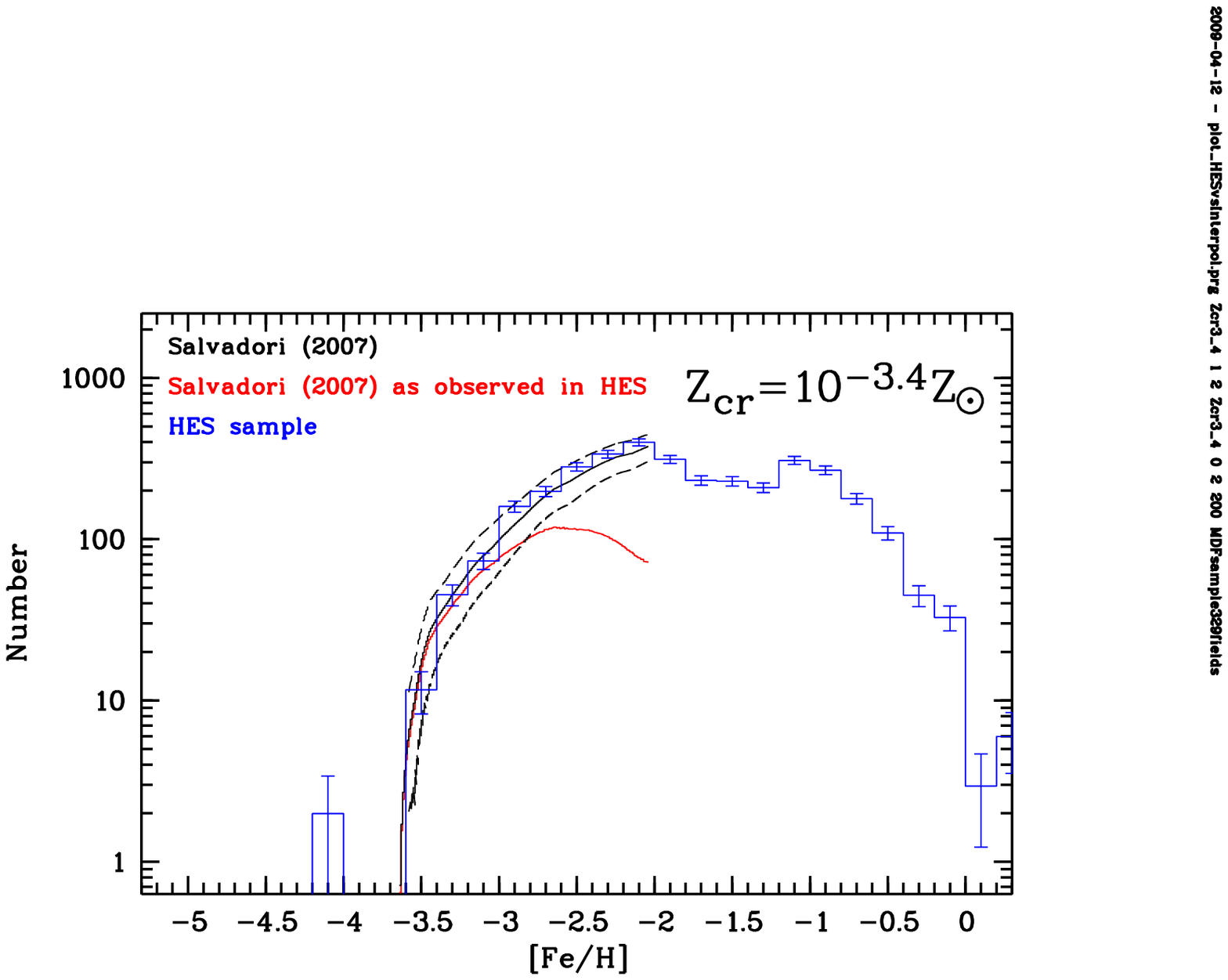}
  \caption{\label{Fig:MDF_Salvadori} Comparison of the MDF constructed from
      the HES sample (histogram) with models of \citet{Salvadorietal:2007}
      with different critical metallicities $Z_{\mathrm cr}$ (black and grey
      lines). The region between the dashed lines indicates the uncertainty of
      the models due to different hierarchical merger histories of the
      Galaxy.}
\end{figure}

\begin{figure}[htbp]
  \centering
  \includegraphics[clip=true,bb=85 420 485 685,width=8.8cm]{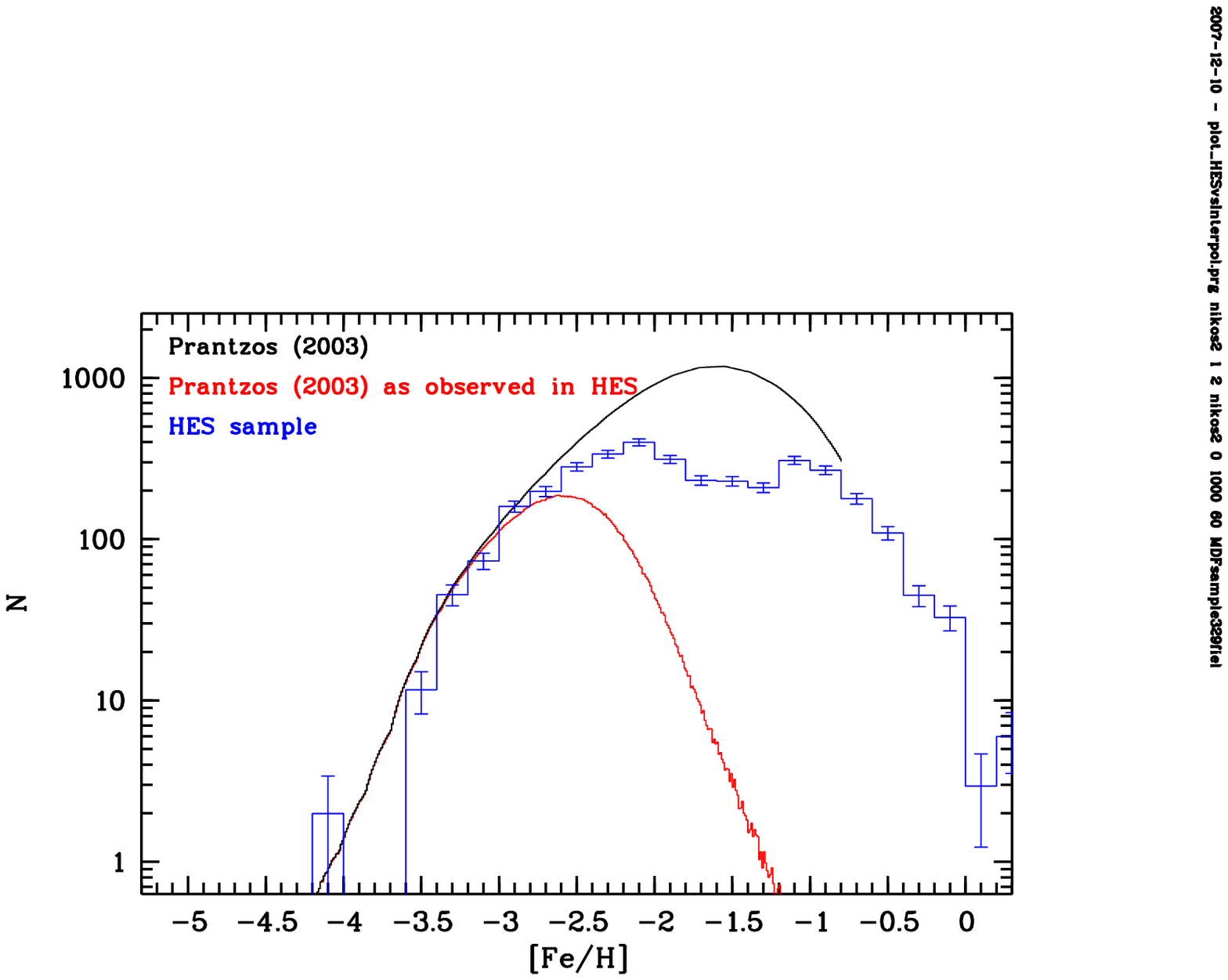}\\
  \includegraphics[clip=true,bb=85 420 485 685,width=8.8cm]{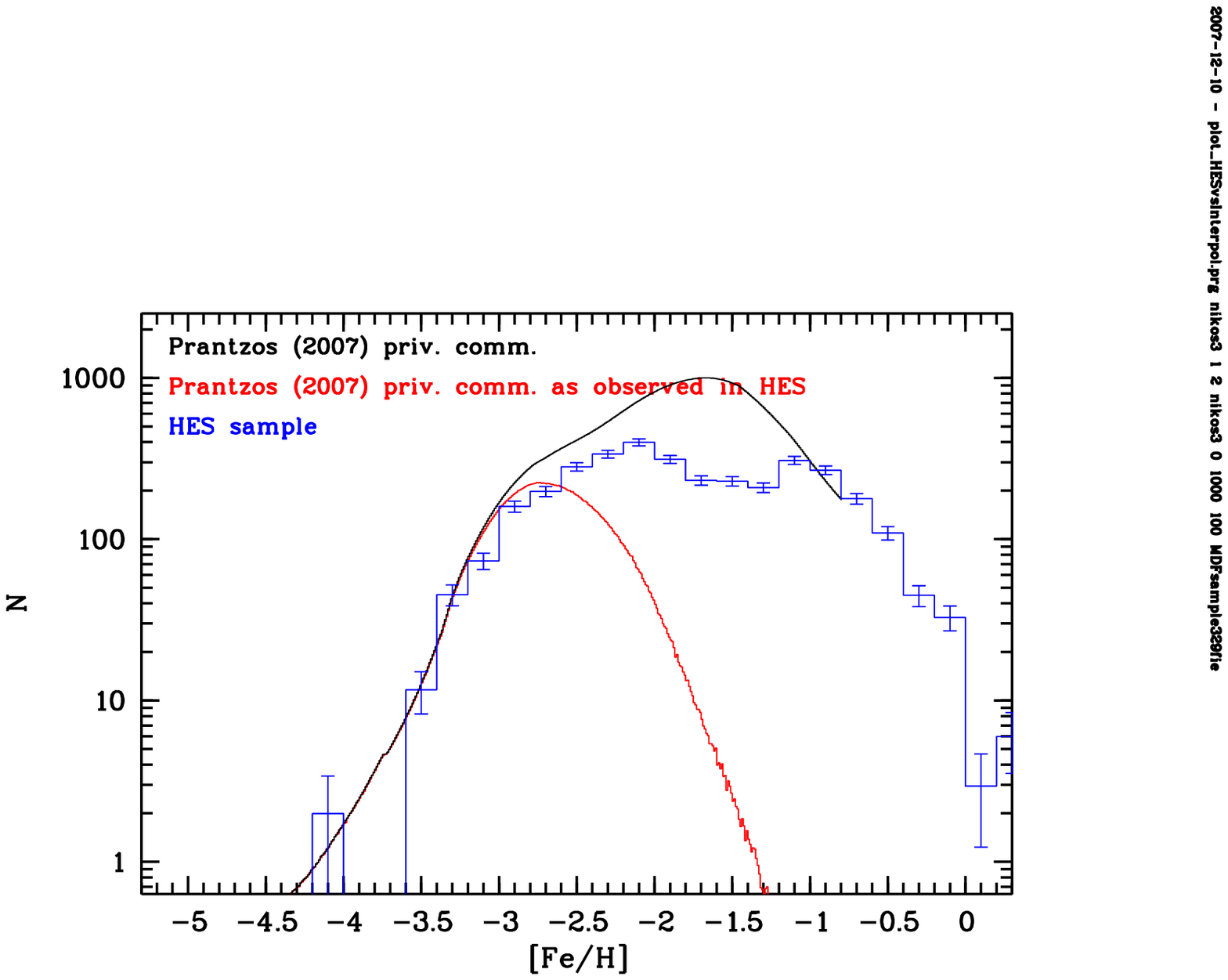}
  \caption{\label{Fig:MDF_Prantzos} Comparison of the MDF constructed from the
      HES sample (histogram) with theoretical predictions (black and grey
      lines). Upper panel: \citet{Prantzos:2003}; lower panel:
      \citet{Prantzos:2007}.}
\end{figure}

\begin{figure}[htbp]
  \centering
  \includegraphics[clip=true,bb=85 420 485 685,width=8.8cm]{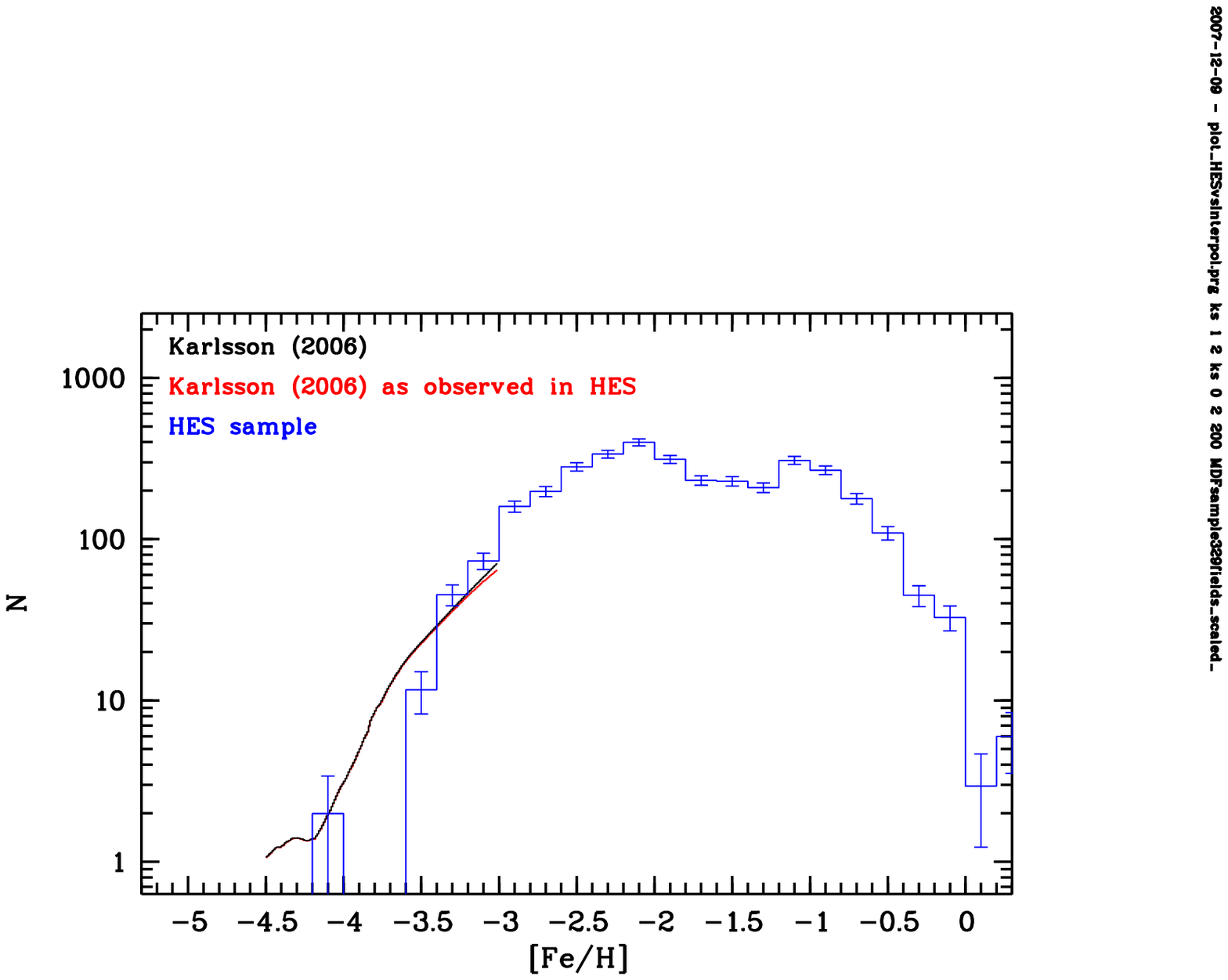}
  \caption{\label{Fig:MDF_Karlsson} Comparison of the halo MDF constructed
    from the HES sample with the MDF predicted by the stochastic
    chemical enrichment model of \citet{Karlsson:2006}. }
\end{figure}

In Fig.~\ref{Fig:MDF_SimpleModel}, we compare the MDF of a Simple Model with
$y_{\mbox{\scriptsize eff}} = -1.7$ with the MDF observed in the HES. The HES
MDF shows an excess of stars in the range $-3.5 < \mathrm{[Fe/H]} < -3.0$.
Alternatively, if the MDF of the Simple Model would be scaled such that it
matches the observed MDF in this range, a large deficit of the number of
observed stars in the range $-3.0 < \mathrm{[Fe/H]} < -2.0$ with respect to
the Simple Model would result. It is also neither possible to reproduce with
the Simple Model the sharp drop of the observed MDF at $\mathrm{[Fe/H]}=-3.6$,
nor the tail at $\mathrm{[Fe/H]}<-3.6$.

\citet{Prantzos:2003} developed a modification of the Simple Model, which
includes early infall, and later outflow of gas; the IRA is also relaxed in
his model. \citet{Prantzos:2007} suggested that since the halo of the Galaxy
has been assembled by merging of a large number of fragments, the MDF of the
Galactic halo can be seen as the sum of the MDFs of these fragments. In his
model, the chemical evolution histories of each of the fragments are still
described by the Simple Model, using the observed mass-metallicity relation of
dwarf galaxies to derive individual effective yields. The halo MDF is then
produced by integrating over a mass function of the fragments determined in
numerical simulations. The MDFs of the models of
\citep{Prantzos:2003,Prantzos:2007} are shown in Fig.~\ref{Fig:MDF_Prantzos}.
Both MDFs match the HES MDF well in the range $-3.5 < \mathrm{[Fe/H]} < -2.5$
and at $\mbox{[Fe/H]}\sim -4.0$, but the sharp drop at $\mathrm{[Fe/H]}\sim
-3.6$ is not predicted by them.

The next set of models that we consider are those of
\citet{Salvadorietal:2007}, who reconstruct the merger tree of the Milky Way
with a semi-analytic approach including a chemical evolution code. A free
parameter in this model is the critical metallicity for low-mass star
formation, $Z_{\mathrm cr}$.  As can be seen in Fig.~\ref{Fig:MDF_Salvadori},
the model with $Z_{\mathrm cr}=10^{-3.4}\,Z_{\odot}$ reproduces the drop of
the observed MDF at $\mathrm{[Fe/H]}\sim -3.6$ rather well. However, the model
predicts that no stars at $\mathrm{[Fe/H]}< -3.6$ should exist, while there
are two such stars in our sample, and for about ten additional stars in this
metallicity range abundance analyses based on high-resolution spectroscopy
have been published (see Tab.~4 of the review of
\citealt{Beers/Christlieb:2005}).

The Salvadori et al. model with $Z_{\mathrm cr}=10^{-4.0}\,Z_{\odot}$ matches
the HES MDF at $\mathrm{[Fe/H]} \sim -4.0$, but disagrees in the range $-4.0 <
\mathrm{[Fe/H]} < -3.6$, where $\sim 30$--$70$ stars are predicted, but none
are present in our sample. The model with $Z_{\mathrm cr}=0$ over-predicts the
number of stars in this metallicity range even more strongly, and it greatly
over-predicts that number of stars at $\mathrm{[Fe/H]} < -4.0$, as already
discussed by \citet{Salvadorietal:2007}.

Finally, we compare in Fig.~\ref{Fig:MDF_Karlsson} the HES MDF with that
predicted by the stochastic chemical enrichment model of
\citet{Karlsson:2006}. While the model matches the HES MDF at $\mathrm{[Fe/H]}
\sim -4.0$, and predicts the tail at $\mathrm{[Fe/H]}<-4.0$ that is known to
exist from additional stars published in the recent literature, the drop of
the observed MDF at $\mathrm{[Fe/H]}\sim-3.6$ is not present in the
theoretical MDF.

To quantify our comparisons of the HES MDF with those predicted by the
theoretical models discussed above, we carried out KS-tests of the null
hypotheses $H_0$ that the HES sample and the individual samples simulated
according to the MDFs of the models were drawn from the same parent
distribution, at a significance level of $\alpha=0.01$. The tests were
restricted to $\mathrm{[Fe/H]}<-3.0$, because we are mainly concerned with the
shape of the low-metallicity tail of the MDF. The result of the tests are that
$H_0$ can not be rejected ($p=0.063$) only in the case of the Salvadori et al.
model for $Z_{\mathrm cr}=10^{-3.4}\,Z_{\odot}$. However, we note that the
statistical test of \citet{Kuiper:1962}, which according to
\cite{PressetalC:1992} is more sensitive than the KS-test to differences at
the ends of the two distributions under comparison (i.e., at the lowest and
highest metallicities), yields $p=6.5\cdot 10^{-4}$; i.e., if this test is
employed, $H_0$ would be rejected at high significance. All other models
considered yielded $p < 10^{-3}$, regardless of which of the two tests were
applied.

\section{Comparison of the halo field star MDF with that of other stellar 
  populations}\label{Sect:GCdSph}

It is of great interest to compare the halo MDF with the MDF found for other
stellar populations, in particular for the system of Galactic globular
clusters (hereafter GCs) and for the stars in dwarf spheroidal (dSph)
galaxies. Since the most metal-poor Galactic GC has $\mbox{[Fe/H]}\sim -2.5$,
we need to establish whether or not there is a real deficit of GCs at lower
Fe-metallicities compared to the halo field.

For a proper comparison of the HES MDF with that of other stellar populations,
it is mandatory that the selection function of the HES, as listed in
Tab.~\ref{Tab:selfrac}, be taken into account. The values in that table can be
used to correct the observed MDF for the selection of metal-poor candidates
employed in the HES. This is particularly important at $\mbox{[Fe/H]} >
-2.5$, where the corrections are large, because typically less than half
of the stars are actually picked up by the HES. Note that this incompleteness
is intended, because the main aim of the search for metal-poor stars with the
HES is to identify stars with $\mbox{[Fe/H]} < -3.0$. Therefore, the
selection of candidate metal-poor stars was designed such that as many stars
at $\mbox{[Fe/H]} > -3.0$ as possible are rejected, while maintaining a
high degree of completeness at $\mbox{[Fe/H]} < -3.0$ (see
\citealt{Christliebetal:2008} for details).

For a star of a given [Fe/H], the corrections are also a function of $B-V$
color, being higher (more likely for a star to be included in the HES) for
redder stars. The variation over the $B-V$ color range of the HES sample can,
in extreme cases at the higher metallicities, correspond to a variation of a
factor of 8 in selection efficiency (see, e.g., the line for
$\mbox{[Fe/H]}=-1.95$ in Tab.~\ref{Tab:selfrac}).

\begin{table}[htbp]
 \centering
 \caption{\label{Tab:cumulativeMDF} Cumulative halo MDF for $\mbox{[Fe/H]} <
   -2.0$ as observed in the HES (column ``Raw''), and corrected for the 
   selection efficiency of the survey (columns 3--5). For details see text.}
  \begin{tabular}{lcccc}\hline\hline
                &          & \multicolumn{3}{c}{$(B-V)_0$ adopted for correction}\\
    \rb{[Fe/H]} & \rb{Raw} & Star & $0.5$ & $1.0$\\\hline
    $-4.30$ & 0.0000 & 0.0000 & 0.0000 & 0.00\rule{0.000ex}{2.3ex}\\
    $-4.25$ & 0.0000 & 0.0000 & 0.0000 & 0.00\\
    $-4.20$ & 0.0000 & 0.0000 & 0.0000 & 0.00\\
    $-4.15$ & 0.0013 & 0.0001 & 0.0003 & 0.00\\
    $-4.10$ & 0.0013 & 0.0001 & 0.0003 & 0.00\\
    $-4.05$ & 0.0013 & 0.0001 & 0.0003 & 0.00\\
    $-4.00$ & 0.0013 & 0.0001 & 0.0003 & 0.00\\
    $-3.95$ & 0.0013 & 0.0001 & 0.0003 & 0.00\\
    $-3.90$ & 0.0013 & 0.0001 & 0.0003 & 0.00\\
    $-3.85$ & 0.0013 & 0.0001 & 0.0003 & 0.00\\
    $-3.80$ & 0.0013 & 0.0001 & 0.0003 & 0.00\\
    $-3.75$ & 0.0013 & 0.0001 & 0.0003 & 0.00\\
    $-3.70$ & 0.0013 & 0.0001 & 0.0003 & 0.00\\
    $-3.65$ & 0.0013 & 0.0001 & 0.0003 & 0.00\\
    $-3.60$ & 0.0013 & 0.0001 & 0.0003 & 0.00\\
    $-3.55$ & 0.0063 & 0.0005 & 0.0014 & 0.00\\
    $-3.50$ & 0.0063 & 0.0005 & 0.0014 & 0.00\\
    $-3.45$ & 0.0094 & 0.0007 & 0.0022 & 0.00\\
    $-3.40$ & 0.0125 & 0.0009 & 0.0029 & 0.00\\
    $-3.35$ & 0.0150 & 0.0011 & 0.0035 & 0.00\\
    $-3.30$ & 0.0201 & 0.0015 & 0.0046 & 0.00\\
    $-3.25$ & 0.0263 & 0.0019 & 0.0060 & 0.00\\
    $-3.20$ & 0.0370 & 0.0028 & 0.0085 & 0.00\\
    $-3.15$ & 0.0489 & 0.0037 & 0.0112 & 0.00\\
    $-3.10$ & 0.0577 & 0.0044 & 0.0132 & 0.00\\
    $-3.05$ & 0.0759 & 0.0058 & 0.0174 & 0.00\\
    $-3.00$ & 0.0828 & 0.0063 & 0.0190 & 0.00\\
    $-2.95$ & 0.1078 & 0.0084 & 0.0250 & 0.01\\
    $-2.90$ & 0.1279 & 0.0102 & 0.0298 & 0.01\\
    $-2.85$ & 0.1567 & 0.0129 & 0.0369 & 0.01\\
    $-2.80$ & 0.1868 & 0.0158 & 0.0445 & 0.01\\
    $-2.75$ & 0.2232 & 0.0196 & 0.0543 & 0.01\\
    $-2.70$ & 0.2589 & 0.0238 & 0.0646 & 0.02\\
    $-2.65$ & 0.2652 & 0.0246 & 0.0665 & 0.02\\
    $-2.60$ & 0.3179 & 0.0319 & 0.0838 & 0.02\\
    $-2.55$ & 0.3348 & 0.0346 & 0.0901 & 0.02\\
    $-2.50$ & 0.3611 & 0.0395 & 0.1010 & 0.03\\
    $-2.45$ & 0.4088 & 0.0498 & 0.1242 & 0.03\\
    $-2.40$ & 0.4815 & 0.0680 & 0.1650 & 0.05\\
    $-2.35$ & 0.5229 & 0.0822 & 0.1927 & 0.06\\
    $-2.30$ & 0.5768 & 0.1037 & 0.2354 & 0.08\\
    $-2.25$ & 0.6213 & 0.1288 & 0.2756 & 0.10\\
    $-2.20$ & 0.6796 & 0.1725 & 0.3413 & 0.14\\
    $-2.15$ & 0.7605 & 0.2657 & 0.4522 & 0.22\\
    $-2.10$ & 0.8031 & 0.3371 & 0.5226 & 0.29\\
    $-2.05$ & 0.8859 & 0.5323 & 0.6851 & 0.49\\
    $-2.00$ & 0.9298 & 0.6759 & 0.7922 & 0.64\\
    $-1.95$ & 1.0000 & 1.0000 & 1.0000 & 1.00\\\hline
  \end{tabular}
\end{table}

\begin{figure}[htbp]
  \centering
%
  \includegraphics[clip=true,bb=55 40 425 496,width=8.8cm]{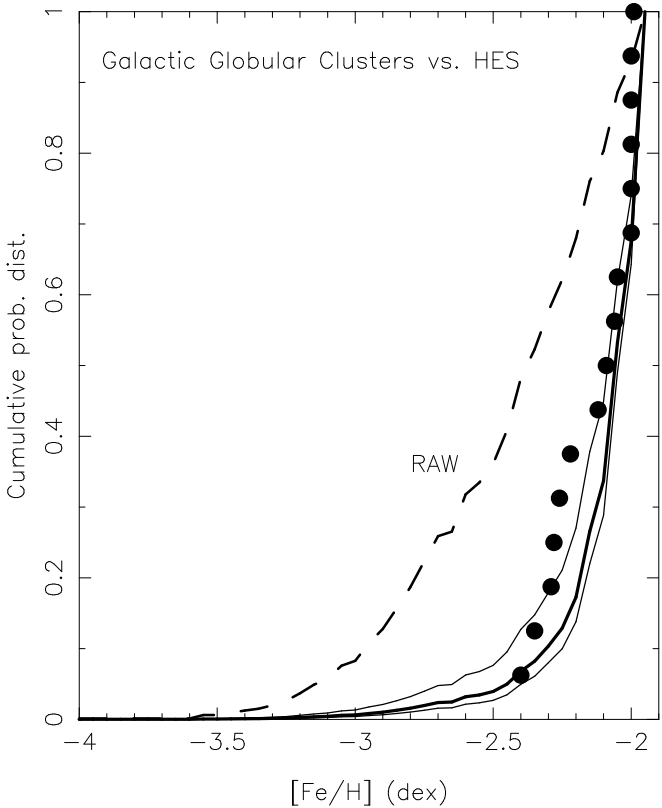}
  \caption{\label{Fig:GC} Cumulative MDF for $\mbox{[Fe/H]}<-2.0$ as observed
    in the HES (dashed line), and with corrections for the HES selection
    efficiency applied (solid lines). Three different ways of applying the
    corrections are shown to illustrate their uncertainty: Multiplication of the
    observed metal-poor star counts with the corrections for $(B-V)_0=1.0$
    (upper solid line), the $(B-V)_0$ color appropriate for each individual
    star in the HES sample (thick, middle line), and the corrections for
    $(B-V)_0=0.5$ (lower solid line). The cumulative MDF of the GCs
    $\mbox{[Fe/H]}<-1.95$ is shown by filled circles. It agrees well with that
    of the halo field stars if the selection efficiency corrections are
    applied.}
\end{figure}

For our comparison with the MDF of the Galactic GCs we adopt the
$\mbox{[Fe/H]}$ values from the current version of the on-line database of
\citet{Harris:1996}. The values for M15 and for NGC~7099 were updated with
small corrections based on detailed abundance analyses carried out by J.~Cohen
and collaborators (Cohen \& Huang, in preparation; Cohen, Melendez \&
Huang, in preparation). The HES is (intentionally) incomplete for
$\mbox{[Fe/H]} > -2.0$, so we only consider the set of GCs with $\mbox{[Fe/H]}
< -1.95$, which contains only 16 clusters. We note that many analyses have
shown that the Galactic GCs exhibit the same behaviour of abundance ratios
(such as the increase of [Ca/Fe] with decreasing [Fe/H]) as the halo stars
(e.g., Fig.~23 of \citealt{KeckpaperIV}) as do the halo stars. Thus, the
conversion between a Ca line index and [Fe/H] adopted by the HES should be
appropriate for Galactic GCs stars as well.

Fig.~\ref{Fig:GC} shows the cumulative MDF from the HES sample and for the
Galactic GC system. The raw MDF and that corrected for the selection
efficiency of the HES, given in Tab.~\ref{Tab:selfrac}, are shown. Note that
the selection efficiency takes into account the uncertainties for [Fe/H] which
result from the uncertainties of the measurement of KP and $(B-V)$ from HES
spectra, which result in $\sigma_{\mathrm{[Fe/H]}}$ between 0.2\,dex and
1.0\,dex, depending on $(B-V)$ colour \citep{Christliebetal:2008}.
Simulations suggest that for a sample with more accurate [Fe/H]
determinations, such as the Galactic GCs, the maximum difference in the
cumulative distribution up to $\mathrm{[Fe/H]} \sim -2.0$ compared to that
given in Tab.~\ref{Tab:selfrac} is small and does not exceed the difference
between the various corrected cumulative MDFs given in columns 3 to 5 of that
table.

The solid, middle line in Fig.~\ref{Fig:GC} corresponds to the case where
corrections according to the dereddened $B-V$ color of each individual star of
the HES sample have been applied. Since these corrections are themselves
uncertain, two other variants are shown in this figure, and listed in
Tab.~\ref{Tab:cumulativeMDF}, to indicate the potential impact of the choice
of $B-V$ color on the corrections.  The first adopts the corrections for the
bluest $B-V$ color of Tab.~\ref{Tab:selfrac}, which are always the smallest,
while the the second uses that of the reddest $B-V$ color of
Tab.~\ref{Tab:selfrac}, which are always the largest.

Fig.~\ref{Fig:GC} shows that once the selection efficiency corrections given
Tab.~\ref{Tab:selfrac} are applied, the halo field star MDF we deduce here is
a good match to that of the Galactic GCs. Instead of expecting roughly 10\,\%
of the sample covering the range $\mbox{[Fe/H]}< -1.95$ to have
$\mbox{[Fe/H]} < -3.0$, we expect only $\sim 2$\,\% to be this metal deficient,
when the selection efficiency for the HES is taken into account. At
$\mbox{[Fe/H]} < -2.5$, the expected fraction decreases from 50\,\% to
8\,\%. Thus, the absence of any GC more metal-poor than $-2.5$\,dex among
a sample of 16 clusters at $\mbox{[Fe/H]} < -1.95$ is not suprising.

A similar situation holds for the stellar population in the dSph satellites of
the Galaxy. It has been widely claimed (see, e.g. the review by
\citealt{Geisleretal:2007}) that these dSph stellar populations show a
significant lack of stars with Fe-metallicity at $\mbox{[Fe/H]}<-3.0$.  For
example, \citet{Helmietal:2006} make this claim for the four systems for which
they assembled the necessary data; i.e., Carina, Fornax, Sculptor and Sextans.

Abundances are now available for large samples of stars in the nearest dSph
galaxies. We concentrate here on those where there is little or no evidence
for recent star formation and for which suitable samples are available. There
are two additional issues that arise in a comparison of the stellar population
of the dSph galaxies with the Galactic halo MDF. The first is that these
metallicities are derived from line indices which measure the strength of the
Ca infrared triplet (CaT) in moderate-resolution spectra. The conversion from
a Ca abundance to a Fe abundance is a crucial issue, since the dSph stellar
population clearly shows a different trend of [Ca/Fe] versus [Fe/H] than does
the Galactic halo (see, e.g., \citealt{Geisleretal:2005} or
\citealt{Monacoetal:2007}), with [Ca/Fe] being smaller at a given
Fe-metallicity in dSph galaxies as compared to GCs and the halo field.  The
second is how the sample to be observed spectroscopically in the dSph is
selected. If, e.g., an equal number of stars in each color bin is chosen to
probe the full range of color across the upper RGB in a dSph, the sample may
be biased in metallicity, because the position of the upper RGB in the
color-magnitude diagram depends on [Fe/H]. Instead, a representative subset of
stars reflecting the color distribution of the stars on the RGB should be
chosen.


\begin{figure*}[htbp]
  \centering

%
  \includegraphics[clip=true,bb=35 15 425 495,width=8.8cm]{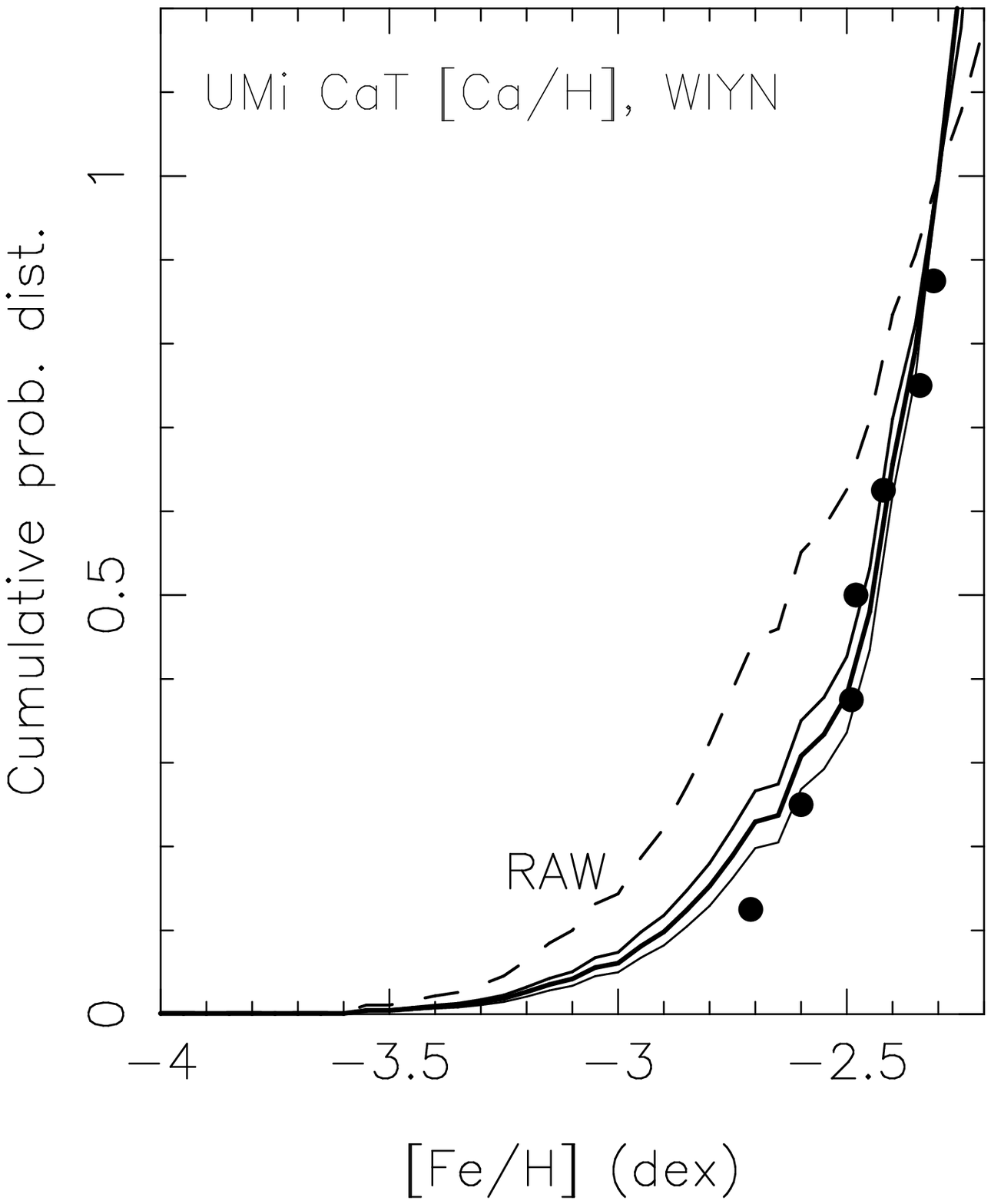}
  \includegraphics[clip=true,bb=35 40 425 495,width=8.8cm]{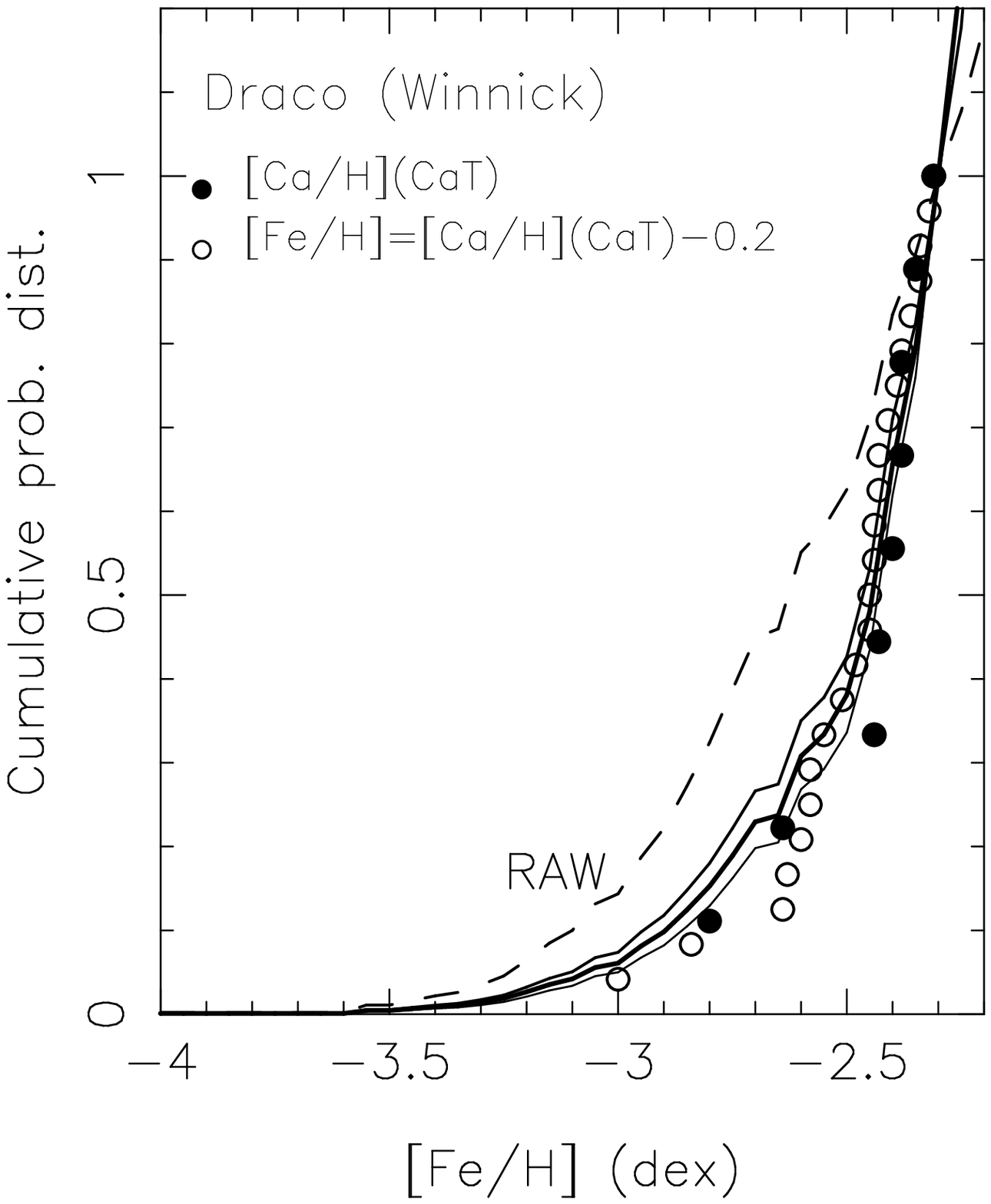}\\[0.5ex]
  \includegraphics[clip=true,bb=35 15 425 495,width=8.8cm]{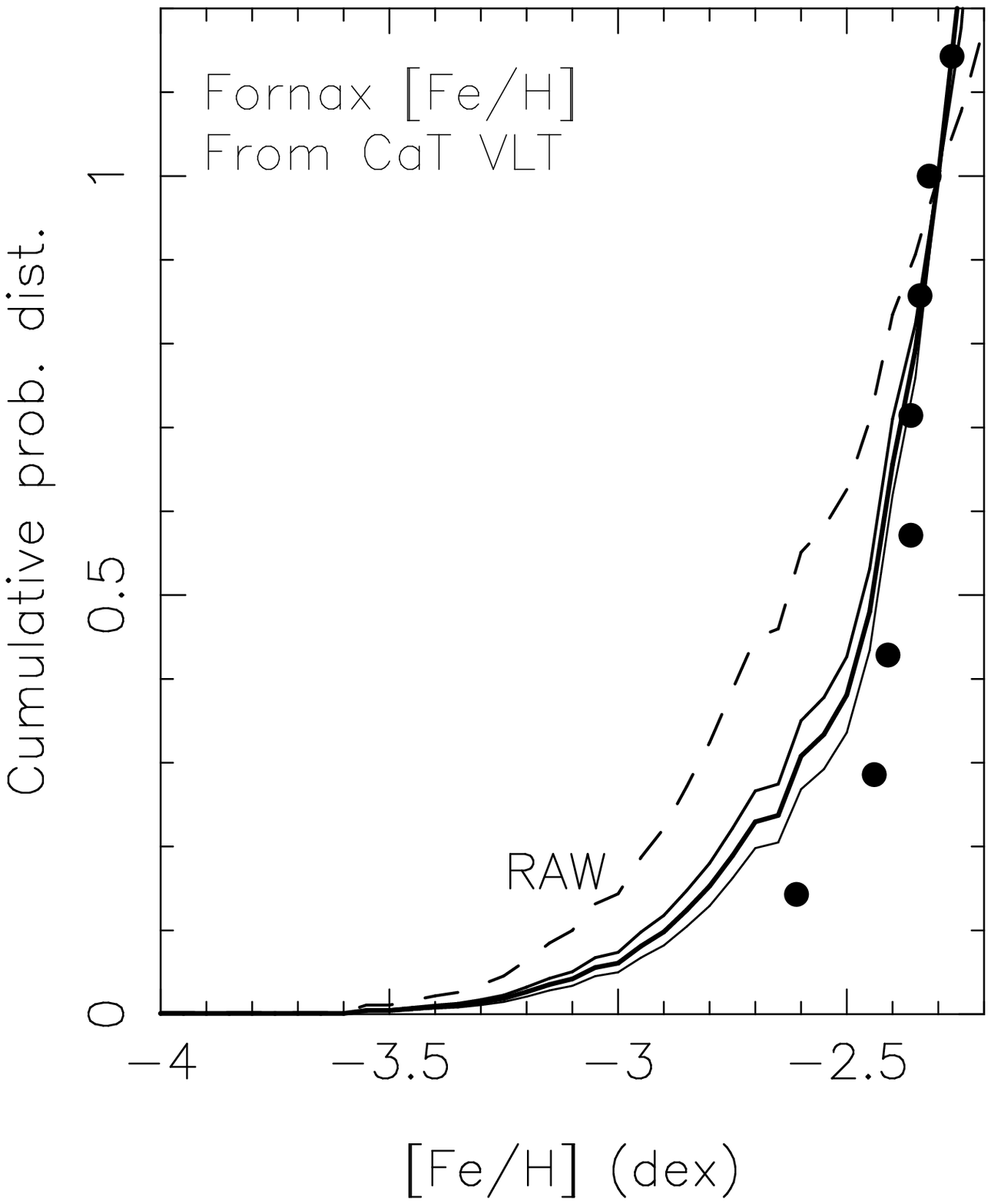}
  \includegraphics[clip=true,bb=55 40 425 495,width=8.8cm]{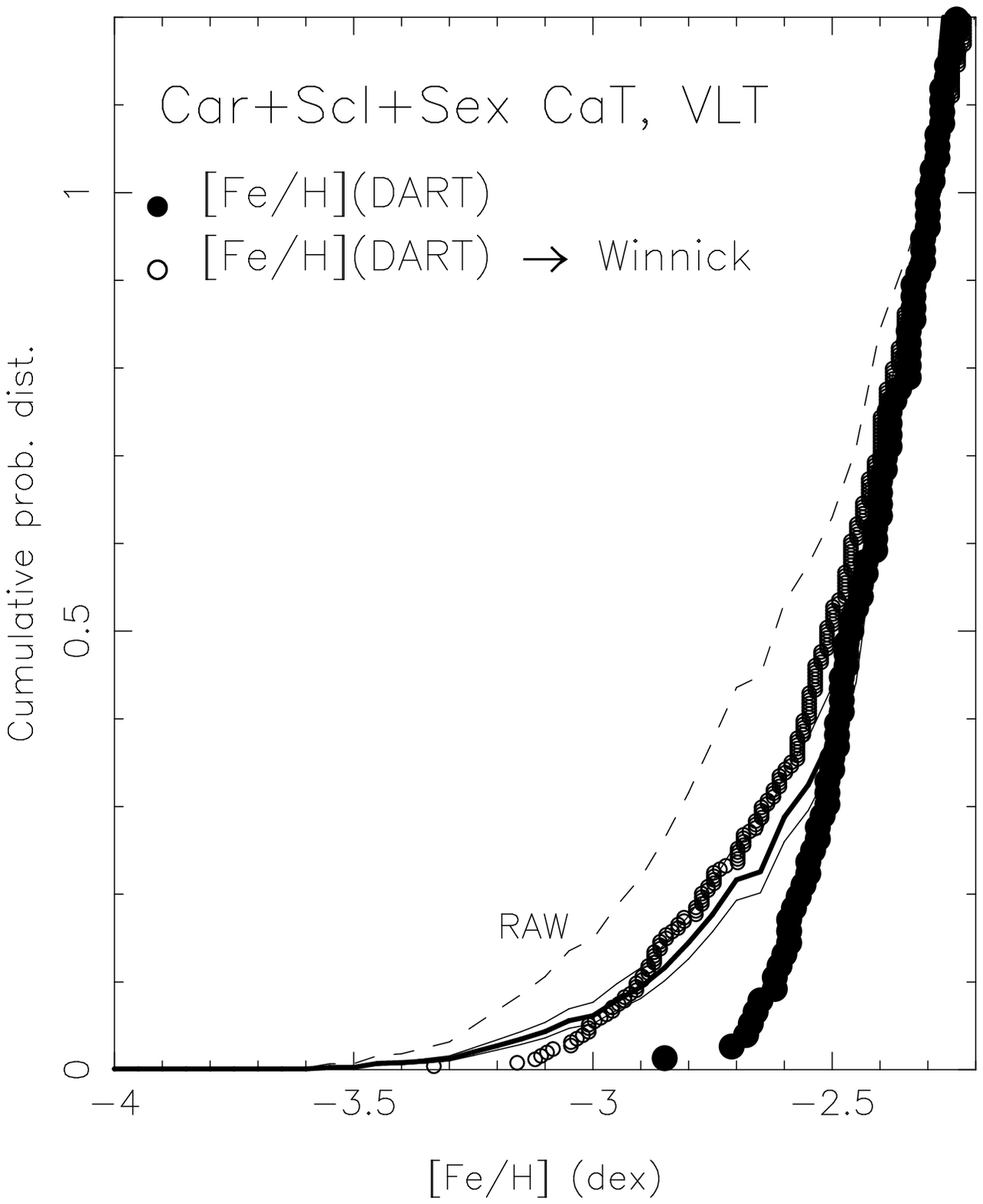}
  \caption{\label{Fig:dSph} 
    Cumulative MDF for $\mathrm{\bf [Fe/H]}\mathbf{<-2.3}$ as observed in the
    HES (dashed line), and with corrections for the HES selection efficiency
    applied (solid lines; see the caption of Fig.~\ref{Fig:GC} and the text
    for a detailed explanation). The filled circles indicate the cumulative
    MDFs of Ursa Minor (upper left) and Draco (upper right), using the [Ca/H]
    values from \citet{Winnick:2003}; for Draco we also show an estimate of
    the MDF for [Fe/H], assuming that $\mathrm{[Fe/H]} = \mathrm{[Ca/H]}-0.2$
    (open circles). In the lower left panel we show a comparison of the
    cumulative MDF for Fornax from \citet{Battagliaetal:2006}. In the lower
    right panel the combined cumulative MDFs of Carina, Sextans, and Sculptor
    based on [Fe/H] values determined in the DART project (filled circles;
    this data was kindly provided by the DART team) is compared to that of the
    HES. The result adopting the calibration of \citet{Winnick:2003} instead
    is shown as the open circles.}
\end{figure*}

%
%

Bearing these caveats in mind, we have constructed the cumulative MDF for
several dSph galaxies. Given the larger samples of very low-metallicity stars
in these galaxies as compared to the limited number of such low metallicity
Galactic GCs, we have normalized the dSph MDFs to $\mathrm{[Fe/H]}=-2.3$.  The
selection efficiency of the HES over this lower metallicity range will be
higher (i.e. closer to 1.0) and not vary as much than is the case over the
regime we needed to consider for the Galactic GCs.  The MDFs for Draco and for
Ursa Minor (top row in Fig.~\ref{Fig:dSph}) are based on the database of
\citet{Winnick:2003}. She measured CaT line strengths from spectra obtained
with the multi-fiber instrument Hydra at the WIYN telescope. Her sample is
selected from radial-velocity members with no metallicity bias.  Winnick
calibrates a relation between both [Ca/H] and [Fe/H] and CaT from observations
of GC giants, making no attempt to take into account the difference in the
behavior of [Ca/Fe] with [Fe/H] in these two stellar populations. We use her
[Ca/H](CaT) values in the figure (solid points).  The detailed abundance
analyses of \citet{Cohen/Huang:2009} suggest that $\mathrm{[Fe/H]}\sim
\mathrm{[Ca/H](CaT,Winnick)}-0.2$ for Draco giants; this yields the open
circles in the Draco panel of Fig.~\ref{Fig:dSph}. The (constant) offset
arises primarily from the lower [Ca/Fe] seen among dSph giants as compared to
GC giants, and includes any error in the calibration adopted by
\citet{Winnick:2003}. The MDF for Draco, with 24 stars at $\mbox{[Fe/H]} <
-2.3$ (nine of which have $\mathrm{[Ca/H]} < -2.3$), and for UMi, agree
reasonably well with the halo Fe-MDF, once the selection efficiency of the HES
is taken into consideration.

For the Fornax dSph we use the VLT/FLAMES$+$GIRAFFE survey of
\citet{Battagliaetal:2006} (their Tab.~4). The DART team in this paper
converted their measurements of the strength of the infrared Ca triplet into
Fe-metallicities (filled circles) using the relation established by
\citet{Rutledgeetal:1997}, which was calibrated using globular cluster giants.
Battaglia advises (priv. comm.) that their sample should be unbiased with
respect to metallicity.  Although there are only 7 Fornax stars in the sample
of \citet{Battagliaetal:2006} with $\mathrm{[Fe/H]} < -2.3$, the left panel of
Fig.~\ref{Fig:dSph} shows that the Fornax dSph is clearly deficient in such
low metallicity stars relative to the MDF of the halo field stars when the HES
raw counts are used. Once the selection efficiencies are folded in, the Fornax
cumulative MDF at $\mathrm{[Fe/H]}<-2.3$ is much closer to that of the
Galactic halo field stars as inferred from the HES.

The lower right panel of Fig.~\ref{Fig:dSph} shows the cumulative MDF from the
combined DART sample for the Carina, Sextans, and Sculptor dSph galaxies, with
[Fe/H] values kindly supplied by the DART project (filled circles). This
yields a total sample of 76 stars with $\mathrm{[Fe/H]} < -2.3$. Again, once
the selection efficiencies are folded in, the cumulative MDF for these three
dSph galaxies at $\mathrm{[Fe/H]}<-2.3$ is much closer to that of the Galactic
halo field stars as inferred from the HES than when the raw HES counts are
used, but there still appears to be a deficit of stars in the combined dSph
MDF at the lowest metallicity compared to the HES MDF.

\citet{Battagliaetal:2008} discussed the accuracy of their conversion between
Ca triplet line strength and [Fe/H], given the difference in the behavior of
[Ca/Fe] with Fe-metallicity between GCs and dSph populations.  Using a
comparison of high-dispersion abundance analyses with their results from CaT
measurements for a limited sample of dSph giants, they conclude that their
Fe-metallicities are robust to within $\pm 0.2$\,dex.  However, as pointed out
by \citet{Cohen/Huang:2009}, there are substantial differences between the
calibration adopted by the VLT DART project and that of \cite{Winnick:2003},
which suggest that the DART project metallicities are too high for
$\mathrm{[Fe/H]} < -2.0$. Hence, we converted the DART [Fe/H] values to those
that would have been inferred using the CaT calibration to [Fe/H] of
\citet{Winnick:2003}, combining Eq.~13 of \citet{Battagliaetal:2008} with
Eq.~3.5 of \citet{Winnick:2003}. The results are indicated by the open
circles.  The application of the CaT calibration of \citet{Winnick:2003} to
the DART data produces a better agreement with the HES Galactic halo MDF.
While the CaT technique appears to be valid even at $\mathrm{[Fe/H]} < -2.5$
\citep{Starkenburg:2009}, the metallicity calibration needs to be improved in
this [Fe/H] range. It is clear from the above that this issue is crucial in
constructing a MDF. Efforts to validate and improve the calibration are
currently underway by \citet{Starkenburg:2009} and others.

We thus find that the MDF of the Galactic halo field stars, as derived from
the HES, agree reasonably well with that of the Galactic globular cluster
system and of the stellar population of the nearest dSph satellites of the
Galaxy, when the calibration for converting infrared Ca triplet line strengths
into [Fe/H] of \cite{Winnick:2003} is adopted. This holds over the range
$\mathrm{[Fe/H]} < -2.3$, after the selection efficiency corrections to the
apparent MDF from the HES have been applied. However, $\chi^2$-tests reveal
that the differences between the halo MDF and the MDFs of the GC system and
the dwarf satellites are still highly significant. If the original DART
calibrations and [Fe/H] are valid, adding the HES selection efficiency
corrections considerably improves the agreement in deduced MDF of the dSph
galaxies with the Galactic halo field stars, but still leaves a problem at the
lowest metallicities.

Recently \citet{Kirbyetal:2008,Kirbyetal:2009} developed a spectral-synthesis
technique that does not use the CaT at all. They found 15 stars with
$\mathrm{[Fe/H]} < -3.0$ in seven of the ultra-faint dSph galaxies recently
discovered by the SDSS. Since all these very low luminosity galaxies have mean
[Fe/H] values of $-1.9$\,dex or lower, this is perhaps not surprising.
\citet{Cohen/Huang:2009} have obtained high resolution spectra of a sample of
stars in the Draco dSph, one of the more luminous of the dSph satellites of
the Galaxy, and found one star with $\mathrm{[Fe/H]} < -3.0$ in that dSph, in
addition to a Draco giant at $\mathrm{[Fe/H]}=-2.97$ discovered earlier by
\citet{Shetroneetal:1998}. In Sculptor, one star with $\mathrm{[Fe/H]} < -3.5$
has recently been identified based on high-resolution spectroscopy (Frebel
2009, priv. comm.).  Finally, in a sample of 16 radial velocity members of the
Bootes~I dSph, \citet{Norrisetal:2008} have reported a giant with
$\mathrm{[Fe/H]} = -3.4$, based of measurements of the Ca~II~K line.
Follow-up, high-resolution, high signal-to-noise observations with VLT/UVES
confirm the result (Norris 2009, priv.comm.). Thus, extremely metal-poor stars
are present, albeit in small numbers, in both the ultra-faint and classical
dSph satellites of the Galaxy.

\section{Discussion and conclusions}\label{Sect:Conclusions}

In Sect.~\ref{Sect:TheoryObservations} we have shown that a reasonable
agreement with the overall shape of the HES MDF can be obtained for
$\mbox{[Fe/H]} > -3.6$ by most models of Galactic chemical evolution, but only
the model of Salvadori et al. with $Z_{\mathrm cr}=10^{-3.4}\,Z_{\odot}$
reproduces the the sharp drop at $\mbox{[Fe/H]} \sim -3.6$ seen in the HES
MDF. The lack of stars at $\mbox{[Fe/H]} < -3.6$ is highly significant: The
models typically predict that about ten such stars should be present in the
HES sample, while only two are found. The significance of this discrepancy is
reflected in the low probabilities for the MDFs predicted by the models and
the HES MDF having the same parent distribution, as determined by KS-tests.
It remains to be investigated whether the drop can be reproduced by modifying
some of the assumptions of the models, or by adding further ingredients.

The HES sample discussed in this paper contains no objects with
$\mathrm{[Fe/H]} < -4.2$, but considering the abundance analyses of three
additional stars in this metallicity range published in the recent literature,
it is obvious that it exists. However, a thorough and quantitative comparison
with theoretical MDFs has to await larger statistically complete and unbiased
samples which include more stars with $\mathrm{[Fe/H]} < -4.0$. Such samples
will become available through new, deeper surveys for metal-poor stars that
will commence in the near future; in particular, the Southern Sky Survey
\citep{Kelleretal:2007} and a survey to be conducted with the Chinese 4\,m
Large sky Area Multi-Object fiber Spectroscopic Telescope (LAMOST;
\citealt{Zhaoetal:2006}).

In the $\Lambda$CDM picture, the Galactic halo was largely built out of
disrupted satellite galaxies. If stars had already formed within them at the
time of accretion, then the MDF of the Galactic halo and of the existing dSph
galaxies should agree at the metal-poor end with regard to the presence of a
weak tail of stars with $\mbox{[Fe/H]} < -3.0$. It is thus encouraging for the
$\Lambda$CDM scenario that our analysis shows better agreement between the
halo MDF and that of the dSph galaxies than claimed by \citet{Helmietal:2006}.
However, even if this were not the case, it would not necessarily be a strong
contradiction to the $\Lambda$CDM scenario. According to the semi-analytical
models of \citet{Salvadorietal:2008} and \citet{Salvadori/Ferrara:2009}, the
MDFs of dSph galaxies can differ quite significantly from each other,
depending on their individual enrichment histories. Hence their MDFs can also
be different from that of the Galactic halo. An important question remaining
to be answered is how the elemental-abundance ratios of the dSph stars at
$\mbox{[Fe/H]} < -3.0$ compare with those of the Galactic halo stars.

Since the HES and the HK survey are in-situ surveys that predominantly sample
the inner-halo population of the Galaxy (with $R < 15$\,kpc), it is mandatory
to consider the possibility that the (for now, poorly studied) outer-halo
population of the Galaxy may indeed contain significant numbers of stars with
$\mathrm{[Fe/H]} < -3.6$, as might be indicated by the shift of the peak
metallicity of the other-halo stars studied by \citet{Carolloetal:2007} to
$\mathrm{[Fe/H]} = -2.2$, a factor of four lower than the peak metallicity of
inner-halo stars. This possibility is being actively pursued by
high-resolution spectroscopic follow-up of stars that are likely to be members
of the outer-halo population, based on their kinematics, by a number of
groups.

\begin{acknowledgements}
  
  We thank T. Karlsson, N. Prantzos, and S. Salvadori for providing us with
  electronic versions of published theoretical MDFs, and for enlightening
  discussions. Valuable comments on an earlier version of this paper by S.
  Ryan are gratefully acknowledged. We thank the DART collaboration for
  providing us with unpublished metallicities of stars in Carina, Sextans and
  Sculptor. N.C. and D.R. acknowledge financial support from Deutsche
  Forschungsgemeinschaft through grants Ch~214/3 and Re~353/44. N.C.  is also
  supported by the Knut and Alice Wallenberg Foundation. J.G.C. is grateful to
  NSF grant AST-0507219 for partial support. T.C.B. acknowledges partial
  funding for this work from grants AST~04-06784, AST~06-07154, AST~07-07776,
  PHY~02-16873, and PHY~08-226498: Physics Frontier Center/Joint Institute for
  Nuclear Astrophysics (JINA), all awarded by the US National Science
  Foundation. M.S.B. and J.E.N. acknowledge support from the Australian
  Research Council under grants DP0342613 and DP0663562. A.F.  acknowledges
  support from the W.J.~McDonald Fellowship of the McDonald Observatory.
  P.S.B is a Royal Swedish Academy of Sciences Research Fellow supported by a
  grant from the Knut and Alice Wallenberg Foundation. P.S.B also acknowledges
  the support of the Swedish Research Council.

\end{acknowledgements}

%
\bibliographystyle{aa}
\bibliography{datanaly,HES,mphs,ncastro,ncpublications,photometry,quasar,TimsRefs}

\end{document}